\documentclass[a4paper,preprintnumbers,amsmath,amssymb,twocolumn,10pt]{revtex4-2}

\begin{document}

\title{Anomaly non-renormalization, lattice QFT
and universality of transport coefficients}
 
\author{Vieri Mastropietro}

\address{University of Milan, Department of Mathematics, via C. Saldini 50, 20129, Milan, Italy}

\begin{abstract} 
Recently new methods have been introduced to investigate
the non-renormalization properties of the anomalies
at a non perturbative level and in presence of a lattice.
The issue is relevant in a number of problems ranging
from the anomaly-free construction of chiral lattice gauge
theory with large cut-off to the universality
properties observed in transport coefficients in condensed
matter systems. A review of main results and future perspectives
is presented.
\end{abstract} 
\maketitle









\newcount\driver
\newcount\bozza

\font\cs=cmcsc10 scaled\magstep1
\font\ottorm=cmr8 scaled\magstep1 \font\msxtw=msbm10
scaled\magstep1                  \font\euftw=eufm10
scaled\magstep1 \font\msytw=msbm10 scaled\magstep1
\font\msytww=msbm8 scaled\magstep1 \font\msytwww=msbm7
scaled\magstep1                 \font\indbf=cmbx10 scaled\magstep2
\font\grbold=cmmib10 scaled\magstep1
\font\amit=cmmi7 \def\sf{\textfont1=\amit} \font\bigtenrm=cmr10
scaled \magstep2               \font\bigteni=cmmi10 scaled
\magstep1

{\count255=\time\divide\count255 by 60
\xdef\hourmin{\number\count255}
        \multiply\count255 by-60\advance\count255 by\time
   \xdef\hourmin{\hourmin:\ifnum\count255<10 0\fi\the\count255}}

\let\a=\alpha \let\b=\beta    \let\g=\gamma     \let\d=\delta     \let\e=\varepsilon
\let\z=\zeta  \let\h=\eta     \let\th=\vartheta \let\k=\kappa     \let\l=\lambda
\let\m=\mu    \let\n=\nu      \let\x=\xi        \let\p=\pi        \let\r=\rho
\let\s=\sigma \let\t=\tau     \let\f=\varphi    \let\ph=\varphi   \let\c=\chi
\let\ps=\psi  \let\y=\upsilon \let\o=\omega     \let\si=\varsigma
\let\G=\Gamma \let\D=\Delta   \let\Th=\Theta    \let\L=\Lambda    \let\X=\Xi
\let\P=\Pi    \let\Si=\Sigma  \let\F=\Phi       \let\Ps=\Psi
\let\O=\Omega \let\Y=\Upsilon

\def\PP{{\cal P}}\def\EE{{\cal E}}\def\MM{{\cal M}}\def\VV{{\cal V}}
\def\FF{{\cal F}}\def\HH{{\cal H}}\def\WW{{\cal W}}
\def\TT{{\cal T}}\def\NN{{\cal N}}\def\BB{{\cal B}}\def\ZZ{{\cal Z}}
\def\RR{{\cal R}}\def\LL{{\cal L}}\def\JJ{{\cal J}}\def\QQ{{\cal Q}}
\def\DD{{\cal D}}\def\AA{{\cal A}}\def\GG{{\cal G}}\def\SS{{\cal S}}
\def\OO{{\cal O}}\def\XXX{{\bf X}}\def\YYY{{\bf Y}}\def\WWW{{\bf W}}
\def\KK{{\cal K}}

\def\ggg{{\bf g}}\def\fff{{\bf f}}\def\ff{{\bf f}}

\def\pp{{\bf p}}\def\qq{{\bf q}}\def\ii{{\bf i}}\def\xx{{\bf x}}
\def\aaa{{\bf a}} \def\bb{{\bf b}} \def\dd{{\bf d}}
\def\yy{{\bf y}}\def\kk{{\bf k}}\def\mm{{\bf m}}\def\nn{{\bf n}}
\def\zz{{\bf z}}\def\uu{{\bf u}}\def\vv{{\bf v}}\def\ww{{\bf w}}
\def\xxi{\hbox{\grbold \char24}} \def\bP{{\bf P}}\def\rr{{\bf r}}
\def\tt{{\bf t}}\def\bT{{\bf T}}

\def\ss{{\underline \sigma}}       \def\oo{{\underline \omega}}
\def\ee{{\underline \varepsilon}}  \def\aa{{\underline \alpha}}
\def\un{{\underline \nu}}          \def\ul{{\underline \lambda}}
\def\um{{\underline \mu}}          \def\ux{{\underline\xx}}
\def\uk{{\underline \kk}}          \def\uq{{\underline\qq}}
\def\uaa{{\underline \aaa}} \def\ub{{\underline\bb}}
\def\uc{{\underlinec}} \def\ud{{\underline\dd}}
\def\up{{\underline\pp}}           \def\ua{{\underline \a}}
\def\ut{{\underline t}}            \def\uxi{{\underline \xi}}
\def\umu{{\underline \m}}          \def\uv{{\underline\vv}}
\def\ue{{\underline \e}}           \def\uy{{\underline\yy}}
\def\uz{{\underline \zz}}
\def\uw{{\underline \ww}}          \def\uo{{\underline \o}}
\def\us{{\underline \s}}           \def\xxx{{\underline \xx}}
\def\kkk{{\underline\kk}}          \def\uuu{{\underline\uu}}
\def\udpr{{\underline\Dpr}}
\def\ggg{\bf g}
\def\uu{\bf u}
\def\III{\hbox{\msytw I}}
\def\MMM{\hbox{\euftw M}}          \def\BBB{\hbox{\euftw B}}
\def\RRR{\hbox{\msytw R}}          \def\rrrr{\hbox{\msytww R}}
\def\rrr{\hbox{\msytwww R}}        

\def\NNN{\hbox{\msytw N}}          \def\nnnn{\hbox{\msytww N}}
\def\nnn{\hbox{\msytwww N}}        \def\ZZZ{\hbox{\msytw Z}}
\def\zzzz{\hbox{\msytww Z}}        \def\zzz{\hbox{\msytwww Z}}
\def\TTT{\hbox{\msytw T}}          \def\tttt{\hbox{\msytww T}}
\def\ttt{\hbox{\msytwww T}}        \def\EE{\hbox{\msytw E}}
\def\eeee{\hbox{\msytww E}}        \def\eee{\hbox{\msytwww E}}

\let\dpr=\partial
\let\circa=\cong
\let\bs=\backslash
\let\==\equiv
\let\txt=\textstyle
\let\io=\infty
\let\0=\noindent

\def\pagina{{\vfill\eject}}
\def\*{{\hfill\break\null\hfill\break}}
\def\bra#1{{\langle#1|}}
\def\ket#1{{|#1\rangle}}
\def\media#1{{\langle#1\rangle}}
\def\ie{\hbox{\it i.e.\ }}
\def\eg{\hbox{\it e.g.\ }}

\def\tilde#1{{\widetilde #1}}

\def\Dpr{\V\dpr\,}
\def\aps{{\it a posteriori}}
\def\lft{\left}
\def\rgt{\right}
\def\der{\hbox{\rm d}}
\def\la{{\langle}}
\def\ra{{\rangle}}
\def\norm#1{{\left|\hskip-.05em\left|#1\right|\hskip-.05em\right|}}
\def\tgl#1{\!\!\not\!#1\hskip1pt}
\def\tende#1{\,\vtop{\ialign{##\crcr\rightarrowfill\crcr
             \noalign{\kern-1pt\nointerlineskip}
             \hskip3.pt${\scriptstyle #1}$\hskip3.pt\crcr}}\,}
\def\otto{\,{\kern-1.truept\leftarrow\kern-5.truept\to\kern-1.truept}\,}
\def\fra#1#2{{#1\over#2}}

\def\sde{{\cs SDe}}
\def\wti{{\cs WTi}}
\def\osa{{\cs OSa}}
\def\ce{{\cs CE}}
\def\rg{{\cs RG}}

\def\lp{{\hskip-1pt:\hskip 0pt}}
\def\rp{{\hskip-1pt :\hskip1pt}}
\def\defi{{\buildrel \;def\; \over =}}
\def\apt{{\;\buildrel apt \over =}\;}
\def\nequiv{\not\equiv}
\def\Tr{\rm Tr}
\def\diam{{\rm diam}}
\def\sgn{\rm sgn}
\def\wt#1{\widetilde{#1}}
\def\wh#1{\widehat{#1}}
\def\hat#1{\wh{#1}}
\def\sqt[#1]#2{\root #1\of {#2}}

\def\ha{{\widehat \a}}\def\hx{{\widehat \x}}\def\hb{{\widehat \b}}
\def\hr{{\widehat \r}}\def\hw{{\widehat w}}\def\hv{{\widehat v}}
\def\hf{{\widehat \f}}\def\hW{{\widehat W}}\def\hH{{\widehat H}}
\def\hB{{\widehat B}}
\def\hK{{\widehat K}} \def\hW{{\widehat W}}\def\hU{{\widehat U}}
\def\hp{{\widehat \ps}}  \def\hF{{\widehat F}}
\def\bp{{\bar \ps}}
\def\hh{{\hat \h}}
\def\jm{{\jmath}}
\def\hJ{{\widehat \jmath}}
\def\hJ{{\widehat J}}
\def\hg{{\widehat g}}
\def\tg{{\tilde g}}
\def\hQ{{\widehat Q}}
\def\hC{{\widehat C}}
\def\hA{{\widehat A}}
\def\hD{{\widehat \D}}
\def\hDD{{\hat \D}}
\def\bl{{\bar \l}}
\def\hG{{\widehat G}}
\def\hS{{\widehat S}}
\def\hR{{\widehat R}}
\def\hM{{\widehat M}}
\def\hN{{\widehat N}}
\def\hn{{\widehat \n}}

\def\PP{{\cal P}}\def\EE{{\cal E}}\def\MM{{\cal M}}\def\VV{{\cal V}}
\def\FF{{\cal F}}\def\HH{{\cal H}}\def\WW{{\cal W}}
\def\TT{{\cal T}}\def\NN{{\cal N}}\def\BB{{\cal B}}\def\ZZ{{\cal Z}}
\def\RR{{\cal R}}\def\LL{{\cal L}}\def\JJ{{\cal J}}\def\QQ{{\cal Q}}
\def\DD{{\cal D}}\def\AA{{\cal A}}\def\GG{{\cal G}}\def\SS{{\cal S}}
\def\OO{{\cal O}}\def\AAA{{\cal A}}

\def\T#1{{#1_{\kern-3pt\lower7pt\hbox{$\widetilde{}$}}\kern3pt}}
\def\VVV#1{{\underline #1}_{\kern-3pt
\lower7pt\hbox{$\widetilde{}$}}\kern3pt\,}
\def\W#1{#1_{\kern-3pt\lower7.5pt\hbox{$\widetilde{}$}}\kern2pt\,}
\def\Re{{\rm Re}\,}\def\Im{{\rm Im}\,}
\def\lis{\overline}\def\tto{\Rightarrow}
\def\etc{{\it etc}} \def\acapo{\hfill\break}
\def\mod{{\rm mod}\,} \def\per{{\rm per}\,} \def\sign{{\rm sign}\,}
\def\indica{\leaders \hbox to 0.5cm{\hss.\hss}\hfill}
\def\guida{\leaders\hbox to 1em{\hss.\hss}\hfill}
\mathchardef\oo= "0521

\def\V#1{{\bf #1}}
\def\pp{{\bf p}}\def\qq{{\bf q}}\def\ii{{\bf i}}\def\xx{{\bf x}}
\def\yy{{\bf y}}\def\kk{{\bf k}}\def\mm{{\bf m}}\def\nn{{\bf n}}
\def\dd{{\bf d}}\def\zz{{\bf z}}\def\uu{{\bf u}}\def\vv{{\bf v}}
\def\xxi{\hbox{\grbold \char24}} \def\bP{{\bf P}}\def\rr{{\bf r}}
\def\tt{{\bf t}} \def\bz{{\bf 0}}
\def\ss{{\underline \sigma}}\def\oo{{\underline \omega}}
\def\xxx{{\underline\xx}}
\let\ciao=\bye
\def\qed{\raise1pt\hbox{\vrule height5pt width5pt depth0pt}}
\def\barf#1{{\tilde \f_{#1}}} \def\tg#1{{\tilde g_{#1}}}
\def\bq{{\bar q}} \def\bh{{\bar h}} \def\bp{{\bar p}} \def\bpp{{\bar \pp}}
\def\Val{{\rm Val}}
\def\indic{\hbox{\raise-2pt \hbox{\indbf 1}}}
\def\bk#1#2{\bar\kk_{#1#2}}
\def\tdh{{\tilde h}}

\def\RRR{\hbox{\msytw R}} \def\rrrr{\hbox{\msytww R}}
\def\rrr{\hbox{\msytwww R}} 
\def\NNN{\hbox{\msytw N}} \def\nnnn{\hbox{\msytww N}}
\def\nnn{\hbox{\msytwww N}} \def\ZZZ{\hbox{\msytw Z}}
\def\zzzz{\hbox{\msytww Z}} \def\zzz{\hbox{\msytwww Z}}
\def\TTT{\hbox{\msytw T}} \def\tttt{\hbox{\msytww T}}
\def\ttt{\hbox{\msytwww T}}

%
\def\ins#1#2#3{\vbox to0pt{\kern-#2 \hbox{\kern#1 #3}\vss}\nointerlineskip}

\newdimen\xshift \newdimen\xwidth \newdimen\yshift

\def\insertplot#1#2#3#4#5#6{%
\xwidth=#1pt \xshift=\hsize \advance\xshift by-\xwidth \divide\xshift by 2%
\begin{figure}[ht]
\vspace{#2pt} \hspace{\xshift}
\begin{minipage}{#1pt}
#3 \ifnum\driver=1 \griglia=#6
\ifnum\griglia=1 \openout13=griglia.ps \write13{gsave .2
setlinewidth} \write13{0 10 #1 {dup 0 moveto #2 lineto } for}
\write13{0 10 #2 {dup 0 exch moveto #1 exch lineto } for}
\write13{stroke} \write13{.5 setlinewidth} \write13{0 50 #1 {dup 0
moveto #2 lineto } for} \write13{0 50 #2 {dup 0 exch moveto #1
exch lineto } for} \write13{stroke grestore} \closeout13
\includegraphics{griglia.ps} \fi
\includegraphics{#4.ps}\fi%
\ifnum\driver=2 \fi
\end{minipage}
\caption{#5}
\end{figure}
}

\def\gtopl{\hbox{\msxtw \char63}}
\def\ltopg{\hbox{\msxtw \char55}}

\newdimen\shift \shift=-1.5truecm
\def\lb#1{%
\ifnum\bozza=1
\label{#1}\rlap{\hbox{\hskip\shift$\scriptstyle#1$}}
\else\label{#1} \fi}

\def\be{\begin{equation}}
\def\ee{\end{equation}}
\def\bea{\begin{eqnarray}}\def\eea{\end{eqnarray}}
\def\bean{\begin{eqnarray*}}\def\eean{\end{eqnarray*}}
\def\bfr{\begin{flushright}}\def\efr{\end{flushright}}
\def\bc{\begin{center}}\def\ec{\end{center}}
\def\bal{\begin{align}}\def\eal{\end{align}}
\def\ba#1{\begin{array}{#1}} \def\ea{\end{array}}
\def\bd{\begin{description}}\def\ed{\end{description}}
\def\bv{\begin{verbatim}}\def\ev{\end{verbatim}}
\def\nn{\nonumber}
\def\Halmos{\hfill\vrule height10pt width4pt depth2pt \par\hbox to \hsize{}}
\def\pref#1{(\ref{#1})}
\def\Dim{{\bf Dim. -\ \ }} \def\Sol{{\bf Soluzione -\ \ }}
\def\virg{\quad,\quad}
\def\bsl{$\backslash$}


%
\def\ins#1#2#3{\vbox to0pt{\kern-#2 \hbox{\kern#1 #3}\vss}\nointerlineskip}

\newdimen\xshift \newdimen\xwidth \newdimen\yshift
\newcount\griglia

\def\insertplot#1#2#3#4#5#6{%
\xwidth=#1pt \xshift=\hsize \advance\xshift by-\xwidth \divide\xshift by 2%
\begin{figure}[ht]
\vspace{#2pt} \hspace{\xshift}
\begin{minipage}{#1pt}
#3 \ifnum\driver=1 \griglia=#6
\ifnum\griglia=1 \openout13=griglia.ps \write13{gsave .2
setlinewidth} \write13{0 10 #1 {dup 0 moveto #2 lineto } for}
\write13{0 10 #2 {dup 0 exch moveto #1 exch lineto } for}
\write13{stroke} \write13{.5 setlinewidth} \write13{0 50 #1 {dup 0
moveto #2 lineto } for} \write13{0 50 #2 {dup 0 exch moveto #1
exch lineto } for} \write13{stroke grestore} \closeout13
\includegraphics{griglia.ps} \fi
\includegraphics{#4.ps}\fi%
\ifnum\driver=2 \fi
\end{minipage}
\caption{#5}
\end{figure}
}

\def\gtopl{\hbox{\msxtw \char63}}
\def\ltopg{\hbox{\msxtw \char55}}

\newdimen\shift \shift=-1.5truecm
\def\lb#1{%
\label{#1}\rlap{\hbox{\hskip\shift$\scriptstyle#1$}}
\else\label{#1} \fi}

\def\be{\begin{equation}}
\def\ee{\end{equation}}
\def\bea{\begin{eqnarray}}\def\eea{\end{eqnarray}}
\def\bean{\begin{eqnarray*}}\def\eean{\end{eqnarray*}}
\def\bfr{\begin{flushright}}\def\efr{\end{flushright}}
\def\bc{\begin{center}}\def\ec{\end{center}}
\def\bal{\begin{align}}\def\eal{\end{align}}
\def\ba#1{\begin{array}{#1}} \def\ea{\end{array}}
\def\bd{\begin{description}}\def\ed{\end{description}}
\def\bv{\begin{verbatim}}\def\ev{\end{verbatim}}
\def\nn{\nonumber}
\def\Halmos{\hfill\vrule height10pt width4pt depth2pt \par\hbox to \hsize{}}
\def\pref#1{(\ref{#1})}
\def\Dim{{\bf Dim. -\ \ }} \def\Sol{{\bf Soluzione -\ \ }}
\def\virg{\quad,\quad}
\def\bsl{$\backslash$}


\driver=1 \bozza=0


\font\msytw=msbm9 scaled\magstep1 \font\msytww=msbm7
scaled\magstep1 \font\msytwww=msbm5 scaled\magstep1
\font\cs=cmcsc10

\let\a=\alpha \let\b=\beta  \let\g=\gamma  \let\d=\delta
\let\e=\varepsilon
\let\z=\zeta  \let\h=\eta   \let\th=\theta \let\k=\kappa \let\l=\lambda
\let\m=\mu    \let\n=\nu    \let\x=\xi     \let\p=\pi    \let\r=\rho
\let\s=\sigma \let\t=\tau   \let\f=\varphi \let\ph=\varphi\let\c=\chi
\let\ps=\Psi  \let\y=\upsilon \let\o=\omega\let\si=\varsigma
\let\G=\Gamma \let\D=\Delta  \let\Th=\Theta\let\L=\Lambda \let\X=\Xi
\let\P=\Pi    \let\Si=\Sigma \let\F=\Phi    \let\Ps=\Psi
\let\O=\Omega \let\Y=\Upsilon

\def\PPP{{\cal P}}\def\EE{{\cal E}}\def\MM{{\cal M}} \def\VV{{\cal V}}
\def\FF{{\cal F}} \def\HHH{{\cal H}}\def\WW{{\cal W}}
\def\TT{{\cal T}}\def\NN{{\cal N}} \def\BBB{{\cal B}}\def\III{{\cal I}}
\def\RR{{\cal R}}\def\LL{{\cal L}} \def\JJ{{\cal J}} \def\OO{{\cal O}}
\def\DD{{\cal D}}\def\AAA{{\cal A}}\def\GG{{\cal G}} \def\SS{{\cal S}}
\def\KK{{\cal K}}\def\UU{{\cal U}} \def\QQ{{\cal Q}} \def\XXX{{\cal X}}

\def\qq{{\bf q}} \def\pp{{\bf p}}
\def\vv{{\bf v}} \def\xx{{\bf x}} \def\yy{{\bf y}} \def\zz{{\bf z}}
\def\aa{{\bf a}}\def\hh{{\bf h}}\def\kk{{\bf k}}
\def\mm{{\bf m}}\def\PP{{\bf P}}

\def\dd{{\boldsymbol{\delta}}}

\def\ddd{\boldsymbol{\d}}
\def\TTTT{\mathbf{T}}

\def\nn{\nonumber}
\def\us{\underset}
\def\os{\overset}

\def\RRR{\hbox{\msytw R}} \def\rrrr{\hbox{\msytww R}}
\def\rrr{\hbox{\msytwww R}}
\def\NNN{\hbox{\msytw N}} \def\nnnn{\hbox{\msytww N}}
\def\nnn{\hbox{\msytwww N}} \def\ZZZ{\hbox{\msytw Z}}
\def\zzzz{\hbox{\msytww Z}} \def\zzz{\hbox{\msytwww Z}}
\def\TTT{\hbox{\msytw T}}


\def\\{\hfill\break}
\def\={:=}
\let\io=\infty
\let\0=\noindent\def\pagina{{\vfill\eject}}
\def\media#1{{\langle#1\rangle}}
\let\dpr=\partial
\def\sign{{\rm sign}}
\def\const{{\rm const}}
\def\tende#1{\,\vtop{\ialign{##\crcr\rightarrowfill\crcr\noalign{\kern-1pt
    \nointerlineskip} \hskip3.pt${\scriptstyle #1}$\hskip3.pt\crcr}}\,}
\def\otto{\,{\kern-1.truept\leftarrow\kern-5.truept\to\kern-1.truept}\,}
\def\defin{{\buildrel def\over=}}
\def\wt{\widetilde}
\def\wh{\widehat}
\def\to{\rightarrow}
\def\la{\left\langle}
\def\ra{\right\rangle}
\def\qed{\hfill\raise1pt\hbox{\vrule height5pt width5pt depth0pt}}
\def\Val{{\rm Val}}
\def\ul#1{{\underline#1}}
\def\lis{\overline}
\def\V#1{{\bf#1}}
\def\be{\begin{equation}}
\def\ee{\end{equation}}
\def\bp{\begin{pmatrix}}
\def\ep{\end{pmatrix}}
\def\bea{\begin{eqnarray}}
\def\eea{\end{eqnarray}}
\def\nn{\nonumber}
\def\pref#1{(\ref{#1})}
\def\ie{{\it i.e.}}
\def\lb{\label}
\def\eg{{\it e.g.}}

\def\Tr{\mathrm{Tr}}
\def\eu{\mathrm{e}}

\newtheorem{lemma}{Lemma}[section]
\newtheorem{theorem}{Theorem}[section]
\newtheorem{cor}{Corollary}[section]
\newtheorem{oss}{Remark}

\section{Introduction} 
.
\subsection{Anomalies and their non-renormalization}

Anomalies are the violation of classical symmetries due
to quantum effects. A paradigmatic example happens in
$QED_4$: the chiral current associated to massless Dirac particles
$j_\m^5$
, conserved at a classical level, verifies 
\cite{Ad}, \cite{AB}
\be \partial_\m j_\m^5={\a\over 4\pi}\e_{\m,\n,\r,\s}
F_{\m,\n} F_{\r,\s}\ee
where the equation has to be understood as an order by
order identity in the perturbation series for correlations.
A similar statement was established 
\cite{GR} in $d=2$
\be\partial_\m j_\m^5={e\over \pi}\e_{\m,\n}\partial_\m
A_\n\ee
A crucial anomaly property is its non-renormalization,
that is the fact that it is an universal quantity independent
on the interaction. The non-renormalization has
the effect that the anomaly can be exactly computed;
this is in sharp contrast with other physical quantities in
QFT which are expressed by series expansions and can
be only computed by truncation at some order. The validity
of the non-renormalization property follows from delicate
cancellations between graphs based on Lorentz
and chiral symmetries \cite{AB}. It also requires regularizations,
like the dimensional one, suitable only in a purely
perturbative context. 

There are derivations of the non-renormalization
avoiding pertubative expansions, like
the one in \cite{F}
, but they essentially neglect higher orders,
see  \cite{AB1}, \cite{K11}. More rigorous derivations have been given 
\cite{S11} 
under the assumption that the gauge fields are classical
smooth fields. Topological explanations hold only with
classical fields \cite{Non}.

\subsection{Anomaly cancellation and universality of
transport}

The anomaly non-renormalization is relevant in a number
of problems ranging from the construction of chiral
lattice gauge theory to the universality properties observed
in transport coefficients.

One of the main features of the Standard Model of elementary
particles is its renormalizability \cite{W}, 
\cite{TH}
, that
is the fact that infinities can be canceled by a choice of
the bare parameters appearing in the action and that
the resulting perturbative expansion is order by order finite.
Such basic property, absent in early theories of weak
forces like the Fermi theory, is obtained, despite the bad
power counting behaviour of massive gauge propagators,
thanks to the Higgs mechanism. The Higgs, however, is
still not sufficient to get renormalizability in the case of
a chiral gauge theory due to the presence of anomalies. The
massive boson propagator is
\be
{1\over k^2+M^2}(\d_{\m\n}+ {k_\m k_\n\over M^2} )
\ee
and the second term, which is not decaying, produces
a non-renormalizable degree of divergence in $d=4$; the
fermion-boson interaction have non-vanishing scaling dimension. In
a non chiral gauge theory with non-vanishing gauge boson mass, Ward Identities associated
to the current conservation $k_\m \hat j_\m=0$
, $\hat j_\m$ the current
in momentum space, ensures that the non decaying
part of the propagator does not contribute to physical
observables. The effective scaling dimension is the
same with or without a boson mass, in particular the
dimension of the fermion-boson interaction is 0. There
is a reduction of the degree of divergence and, for instance, the transition
in QED$_4$ from a massless or a massive photon is
soft; renormalizability is preserved. 

This argument cannot
be applied if the gauge field is coupled to a chiral
current, as in electroweak theory, so that the theory remains
non-renormalizable, unless the anomalies cancel
out. Remarkably it was shown in \cite{1} that the lowest order
contributions to the anomaly vanish if the following
condition is verified
\be
\sum_i Y_i^L-\sum_i  Y_i^R=0\quad \sum_i (Y_i^L)^3-\sum_i  (Y_i^R)^3=0
\label{appa}
\ee
where $i=\n,e,u,d$ runs over a single family, the quarks have three colors and $Y_i^L,Y_i^R$ are the hypercharges of the $L$ and
$R$ particles. The
charges, that at a classical level can take any value, are
constrained by a purely quantum effect, a fact providing
a partial explanation to charge quantization without reference
to grand unification.

The condition \pref{appa} is found from the lowest
order computation of the anomaly, but higher orders
could require extra conditions to vanish, impossible to
satisfy; however the 
non-renormalization property ensures that such
higher order contributions are vanishing, hence renormalizability
is preserved. It should be stressed however that
this is a purely perturbative statement as the series are expected
to be divergent and even not asymptotic, at least
if one considers the electroweak sector alone, due to triviality
of the scalar sector, see eg \cite{12}  .

The anomaly non-renormalization manifests itself
also in a physical context apparently far from high energy
physics, that is the universality phenomena observed in
Condensed Matter. While macroscopic properties of materials
depend in general on all the microscopic details,
there is a class of systems where some transport properties
are universal, in particular they do not depend
on the interaction between conduction electrons. A well
known example is the transverse, or Hall, conductivity
of two-dimensional insulating systems exposed to magnetic
fields, which is is equal to ${n e^2\over h}$ , where $e$ is the
electric charge, $h$ is the Planck constant and n is
an integer. Another example is the optical conductivity
in Graphene which is experimentally found equal to \cite{14} 
\be
\s={e^2\over h}{\pi\over 2}
\ee
In certain cases, as in Hall systems, universality
has a topological origin but in the case of Gr
raphene, for instance, such interpretation is not possible.
The universality properties in transport coefficients
have been related to the non renormalization of anomalies,
see e.g. \cite{13}. Electrons in metals are non relativistic
fermions, described by the Schroedinger and not by
the Dirac equation; nevertheless the interaction with the
lattice can produce an emerging effective description in
terms of an Euclidean QFT models, in particular when
the Fermi surface is point-like. This happens in Hall systems
or Graphene, where there is an emerging description
in terms of massive or massless QED$_3$ \cite{15},\cite{16}, in
Weyl semimetals described by massless QED$_4$ \cite{17}
 and in
one dimensional metals described by QED$_2$ \cite{18}. Due to
this fact, for instance, the universality of Hall conductivity was
connected to a non-renormalization properties of
the emerging QED$_3$ description, see e.g. \cite{19},\cite{20}. Note
however that such explanations are based on cancellations
in series expansions due to exact continuum and
relativistic symmetries, which are violated by lattice effects
and produce finite contributions unless extra cancellations
occur. They are therefore not sufficient to fully
explain universality of transport, which is observed with
very high precision.

\subsection{Effective lattice QFT and Condensed Matter}
The Standard Model, at least if one considers the electroweak
sector only, has presumably a non-perturbative
meaning only as an effective (Euclidean ) QFT theory, as
it contains the Higgs particle which is affected by the triviality
problem, see e.g. \cite{21},\cite{22}.
Therefore, it must be
defined with a finite cut-off, and it is expected to be replaced
by a more fundamental theory at very high energies.
To be the effective QFT applicable to nature, one
needs that:
\begin{enumerate}
\item 
the effect of the ultraviolet cut-off is negligible at the
energy scales of the experiments;
\item
gauge invariance and Ward Identities are valid with
finite cut-off;
\item
the fermionic masses are vanishing or very small with respect
to the coupling in the same units.
\end{enumerate}

The second requirement is not fulfilled by the momentum
regularization typically used in Renormalization
Group (RG) analysis, like the ones in \cite{23}-\cite{25}, based on
tree expansion \cite{26}, and in \cite{27}-\cite{29}, using flow equations
\cite{30}. Ward Identities in these analysis are recovered
only when cut-offs are removed, hence they are suitable
only in a perturbative context and not in an effective
non perturbative one. 
Non-perturbative constructions of
QED \cite{30a}
removing the momentum cut-off assumed that
fermion masses are much greater than charges, so violating
the third requirement.

The use of a lattice cut-off has the merit of preserving
vector Ward Identites and is the more appropriate for the
construction of an effective QFT. A non-perturbative control
cannot however be achieved for any value of the lattice
step (otherwise the continuum limit could be taken).

There is indeed an expected relation between the size
of the cut-off and the renormalizability properties. In
the case of a lattice version of the (non renormalizable)
Fermi theory one can achieve a non-perturbative control
of the theory (see below) for $Ga^2$ smaller than $1$, if $a$ is
the lattice step and $G$ the Fermi constant.
Hence the
range of validity of a lattice Fermi theory is much smaller
that the energy scales of modern experiments. In contrast,
an effective theory for the electroweak gauge theory
could be defined with much higher cut-offs; being renormalizable
with scaling dimension of the boson fermion
interaction equal to zero, one expects in principle a nonperturbative
control for $g^2 \log(aM)^{-1}$ smaller than $1$, if $g$
is the gauge coupling in adimensional units and $M$ is an
energy scale (for instance the mass of gauge bosons); that
is a cut-off exponentially high in the inverse coupling, fulfilling
the first requirement.

However, as discussed above, the renomalizability in
electroweak theory is based on the reduction of scaling
dimension due to the anomaly cancellation and 
the non-renomalization. One needs the
same properties in a non perturbative context with a finite
lattice. There has been an extensive search for a nonperturbative
regulator for lattice chiral gauge theories, directly
ensuring the validity of chiral Ward Identites under
the anomaly cancellation condition, but this is a long
standing unsolved problem 
\cite{B3c}-\cite{B8}
(see e.g. \cite{D1}- \cite{D} for reviews), 
and best results, regarding the $U(1)$ sector,
are only valid order by order \cite{B3}  or with a formal treatment
of the thermodynamic limit \cite{B4} .

In order to get the anomaly cancellation and the reduction
of degree of divergence in lattice electroweak theory,
necessary to get high cut-off, one needs to establish
the cancellations of the anomaly based on the non renormalization
also when symmetries are only emerging and
in presence of a finite lattice. Remarkably, this is the same
kind of problem one encounters for understanding the
universality of transport coefficients, where lattice and
interaction are surely present, like in the conductivity of
graphene \cite{43}-\cite{47}
or in Weyl semimetals \cite{48}-\cite{51}. It
should be stressed that the cancellation of the anomalies
or the universality of transport need to be really exact and
not approximate, either for theoretical or experimental
consistence. Such problems cannot be faced in a perturbative
scheme, due to the lack of convergence and the
complexity of the expansion in presence of a lattice, Indeed,
even if lattice effects are often irrelevant in the RG
sense, they produce finite contributions.

Recently new methods have been introduced allowing
finally to prove the non-renormalization and universality
of anomalies with a finite lattice and excluding nonperturbative
effects, in several situations. This approach
is based on rigorous Renormalization Group (RG), allowing
to express quantities in terms of series converging in
a finite region of the parameters, uniformly in the volume,
and to take into full account the irrelevant terms in
the RG sense. Non renormalization follows from the subtle
interplay of emerging and lattice Ward Identities and
non-perturbative decay bounds for the correlations. A review
of main results and perspectives is here provided

\section{Non-compact Massive $U(1)$ lattice
gauge theory with Wilson fermions}

\subsection{Functional Integral formulation}

We start our analysis from a lattice $U(1)$ vector massive
Gauge model. In $d=4$ its generating functional is
\be
e^{W(J,J^5,\phi)} =
\int P(dA)\int P(d\psi) e^{V_c(\psi)+ V_{e}(A+J,\psi)+B(\psi,J^5,\phi)}\label{ap}  \ee
where
$\L=[0,L]^4\cap a \ZZZ^4$, $\m=0,1,2,3$
$\psi_{x},\bar\psi_x$ are {\it Grassmann variables} with antiperiodic boundary conditions, 
$\g_0= \begin{pmatrix} 0 & I \\ I &0 \end{pmatrix}\quad \g_j= \begin{pmatrix} 0 & i\s_j \\-i\s_j &0 \end{pmatrix}, \quad\g_5=\begin{pmatrix}&I&0\\
          &0&-I\end{pmatrix}$
and $\s_\m^L=(\s_0,i \s) $, $\s_\m^R=(\s_0,-i \s) $. The {\it fermionic integration} is
\be
P(d\psi)={1\over\NN_\psi}  [\prod_{x} d\bar\psi_{x}  d\psi_{x}
] 
e^{-S}\ee
%
with $e_\m$ unit vectors and
\bea
&&S={1\over 2 a}  a^4\sum_x   [  
 \bar\psi_{x}\g_\m   \psi_{x+e_\m a}-\bar\psi_{x+e_\m a}
\g_\m \psi_{x}+\label{no}\\
&&r ( \bar\psi_{x}\psi_{x+e_\m a}+\bar\psi_{x+e_\m a}\psi_{x}
-\bar\psi_{x}\psi_{x}) ]\nn\eea
and
\be
V_c(\psi)=\n a^4 \sum_x \bar\psi_x \psi_x\ee is the mass counterterm.
The  fermionic propagator is
therefore given by
\be
g (x -y)=\int P(d\psi)\psi_x \bar\psi_y={1\over L^4}\sum_k
{e^{i k (x-y)}\over -i \not s(k)+M(k)  }
\ee
with 
\be
\not s(k)=a^{-1}\sum_\m \g_\m \sin a  k_\m\quad M(k)=2 r a^{-1}
\sum_\m \sin^2 k_\m a
\ee
and $k={2\pi\over L}(n+1/2)$. 
It is also convenient to write $\psi=(\psi^-_L, \psi^-_R)$
and $\bar\psi=(\psi^+_R, \psi^+_L)$
where $s=L,R$ denotes the chirality.

If $A_\m(x): \L\to \RRR$
the bosonic integration $
P(dA)$ has propagator 
\be
g^A_{\m,\n}(x,y)=
{1\over L^4}\sum_k  { e^{i k (x-y)}\over |\s|^2+M^2}(\d_{\m,\n}+ {\x \bar\s_\m \s_\n\over (1-\x) |\s|^2+M^2})\label{propo}\ee
with $\s_\m(k)=i(e^{i k_\m a}-1)a^{-1}$. 

The interaction $V_{e}$ is, if $G_{\m}(A+J)=a^{-1}( e^{i e a (A_\m(x)+J_{\m}(x)) }-1)$
\bea
&&V_e={1\over 2 a}  a^4\sum_x   [  
 \bar\psi_{x}\g_\m G^+_\m \psi_{x+e_\m a}-\bar\psi_{x+e_\m a}
\g_\m G^-_\m \psi_{x}+\label{no1}\\
&&r ( \bar\psi_{x} G^+_\m \psi_{x+e_\m a}+\bar\psi_{x+e_\m a}G^-_\m \psi_{x}]\nn\eea
Finally the source term is $B=(\phi,\psi)+\ZZ_5 (J_\m^5,j_\m^5)$ with
\be
j^5_{\m,x}=\ZZ^5 \sum_s \e_s \psi^+_{x,s}\s_\m^s \psi^+_{x,s}\ee
with $\e_{L}=-\e_{R}=1$ is the chiral current. 

Analogously is defined a lattice massive U(1) gauge
theory in $d=2$; in this case $\psi=(\psi^-_+,\psi^-_-)$, 
$\bar\psi=(\psi^+_-,\psi^+_+)$,
$\g_0=\s_1$, $\g_1=\s_2$, $\g_5=\s_3$ 
\be \s_1=\begin{pmatrix}&0&1\\
          &1&0\end{pmatrix}
 \quad
\s_2=\begin{pmatrix}
&0&-i\\ &i&0
\end{pmatrix}
\quad\s_3=\begin{pmatrix}&1&0\\
          &0&-1\end{pmatrix}\ee
Again $s=\pm$ denotes the chirality of the fermions.

From the generating function one can obtain the
correlations; in particular the 2-point function 
$
S(x,y)
={\partial^2 \over \partial \phi^+_{x}\partial \phi^-_{y} }
W(J,J^5,\phi)
|_0$, the vertex functions
\be
\G_\m(z;x,y)={\partial^3 W\over \partial J_{\m,z} \partial\phi_x\partial\bar \phi_y}|_0
\quad \G^5_\m(z,x,y)={\partial^3 W\over \partial J^5_{\m,z} \partial\phi_x\partial\bar \phi_y}|_0\ee 
and the correlations
\bea
&&\Pi_{\m_1,\m_2,...,\m_N}(x_1,x_2,..,x_n) ={\partial^n W\over\partial J_{\m_1,x_1}...\partial J_{\m_n,x_n}}|_0\nn\\
&&\Pi^5_{\m_1,\m_2,...,\m_N}(x_1,x_2,..,x_n) ={\partial^n W\over\partial J^5_{\m_1,x_1}...\partial J_{\m_n,x_n}}|_0
\eea
We denote by $\hat S(k), \hat\G_\m(p,k),\hat\G^5_\m(p,k)$,
$\hat\Pi_{\m_1,\m_2,...,\m_n}(p_1,..,p_{n-1})$
and $\hat\Pi^5_{\m_1,\m_2,...,\m_n}(p_1,..p_{n-1})$
the corresponding Fourier
transforms.

The term proportional to $r$ in (8) is the Wilson term,
necessary to get avoid extra spurious degrees of freedom,
that is extra poles in addition to $k=0$ in the fermionic
propagator. Its presence
breaks the chiral invariance of the theory \cite{52}, 
$\psi^\pm_{x,s}\to e^{\pm i a_s}\psi^\pm_{x,s}$; as a consequence, one has to properly choose
the mass counterterm $\n$ to get an interacting massless
theory, that is such that that $\hat S(k)$ is diverging for $k=0$
in the $L\to\io$ limit. Similarly the breaking of chiral Ward
Identities (see next subsection) requires that the renormalization
$\ZZ_5$ is chosen via the request that the
charge carried by the current and the axial current is the
same, see \cite{AB1}, in the $L\to\io$ limit
\be
\lim_{k,p\to 0}{\hat\G_\m(p,k)\over \hat\G^5_\m(p,k)}=\g_5
\ee
Finally $\x$ is the gauge fixing parameter and $M$ is the boson
mass. The lattice formulation of the vector model (6)
is non-compact, as the boson field is unbounded, but is
still well defined at a non perturbative level, even if in the
$L\to\io$ limit, as shown below.

\subsection{Ward Identities}

Ward Identities (WI) associated to the total current are
valid with a lattice regularization. As the functional intergral
is well defined, we can perform safely the change of
variables, with Jacobian 1,
$\psi_{x}\to \psi_{x} e^{i e\a_x}$ obtaining
\be
W(J,J^5, \phi)=
W(J+d\a ,J^5, e^{i e \a}
 \phi)\nn
\ee
and by differentiating with respect to the $\a_x$ and the external
fields
\bea
&&-i\sum_\m \s_\m(p)
\hat \G_\m(k,p)
= e 
(\hat S(k)-\hat S(k+p))\nn\\
&&\sum_{\m_1} \s_{\m_1}(p_1+..p_{n-1}) \hat\Pi_{\mu_1, \mu_2,..., \m_n}(p_1,..,p_{n-1})=0\label{zippo}
\eea
which express the conservation of the total current.

If $r=\n=0$ also the chiral WI are valid, obtained
via the change of variables $\psi_{x}\to \psi_{x} e^{i e\g_5 \a_x}$; replacing $J_\m$ with $J_\m+ \g_5 \tilde J_\m^5$
in the generating function one gets
\be
\sum_{\m_1} \s_{\m_1}(p_1+..p_{n-1}) \tilde\Pi^5_{\mu_1, \mu_2,..., \m_n}=0
\ee
where $\tilde\Pi^5_{\mu_1, \mu_2,..., \m_n}$
is the derivative with respect to $\tilde J^5_\m$.
Similarly
the WI for the chiral current ensures the nonrenormalization
of the electric charge associated to the
total current and $\ZZ_5=1$. 

In presence of the Wilson term
$r\not=0$ (20) is not verified; in particular one is interested
in $\sum_{\m_1} \s_{\m_1}(p) \Pi^5_{\mu_1, \mu_2}$
in $d=2$ and $\sum_{\m_1} \s_{\m_1}(p_1+p_{2}) 
\tilde\Pi^5_{\mu_1, \mu_2,\m_3}(p_1,p_2)$
in $d=4$ which are the only non vanishing cases in the continuum.

\subsection{Independence on the $\xi$ parameter}

In the case $M\not=0$ gauge invariance in the $A$ fields is broken;
nevertheless the validity of WI ensures that the expectations
of gauge invariant observables
\be
\mathcal{O}(\psi e^{ie \alpha(x)},A+d \alpha_x)
=\mathcal{O}(\psi ,A)\ee
are independent
on  the gauge-fixing parameter $\xi$, that is
\be
\partial_\x 
{\int P(dA)\int P(d\psi) e^{V_c(\psi)+ V_{e}(A+J,\psi)}\mathcal{O}(\psi ,A)
\over \int P(dA)\int P(d\psi) e^{V_c(\psi)+ V_{e}(A+J,\psi)}}=0
\ee
This follows from the fact that, setting
\be
\G(J)=\int P(dA)\int [\prod_x d\psi_x d\bar\psi_x] e^{S_t(A+J,\psi)}\mathcal{O}(\psi ,A+J
)\ee
with $S_t(A+J,\psi)=S(\psi)+V_c(\psi)+ V_{e}(A+J,\psi)$, and 
$G_{\m_1,\m_2,...,\m_n}$
$={\partial^n \G\over\partial J_{\m_1,x_1}...\partial J_{\m_n,x_n}}|_0$ one has 
\be
\sum_{\m_1}\s_{\m_1}\hat G_{\m_1,\m_2,...,\m_n}=0
\ee
Therefore
\bea
&&\partial_\x \G(J)={1\over L^4 }\sum_p 
\partial_\x (\hat g^A(p))^{-1}_{\m,\n} \\
&&\int P(dA) A_{\m,p} A_{\n,-p}
\int \prod d\psi_x d\bar \psi_x e^{S_t(A+J,\psi)}\mathcal{O}(\psi ,A+J)\nn
\eea
from which we get
\bea
&&\partial_\x \G(J)|_0={1\over L^4 }\sum_p \hat g^A_{\r',\m}(p)
\partial_\x (\hat g^A)^{-1}_{\m,\n}\hat g^A_{\n,\r}(p)\hat G_{\r,\r'}=\nn\\
&&{1\over L^4 }\sum_p 
\partial_\x \hat g^A_{\m,\n}(p) \hat G_{\m,\n}=0
\eea

as $\partial_\x \hat g^A_{\m,\n}(p)$
is proportional to $\bar\s_\m \s_\n$. The same is true for any
derivative with respect to $J$. 

The $\xi$-independence ensures
that the averages of (21) can be computed at
$\x=0$, that is in absence of the non decaying part of the
propagator.

\subsection{Lattice chiral anomaly in the non-interacting case}

In the non-interacting case $V_e=V_c=0$ then
$\sum_{\m_1} \s_{\m_1}(p) \hat \Pi^{5,0}_{\mu_1, \mu_2}$
in $d=2$ and $\sum_{\m_1} \s_{\m_1}(p_1+p_2) 
\hat\Pi^{5,0}_{\mu_1, \mu_2,\m_3}$
in $d=4$ have the
same value in the continuum or with a finite lattice, up to
subleading terms in the momentum. 

We show this using a strategy which can be generalized
to the interacting case.We write the lattice fermionic
propagator (10) as
\be
\hat g(k)={\chi(k)\over-i \not k}+r(k)\label{27}
\ee
where $\chi(k)$ is s smooth compact support function vanishing
outside a circle of radius $O(a^{-1})$ excluding the
poles of $\hat S(k)$
except $k=0$ while $|r(k)|\le C$; the first term
corresponds to the propagator in the formal continuum
limit with a momentum regularization while the second is an irrelevant term in the RG terminology, depending on terms which are formally vanishing in the continuum limit.

The decomposition
(27) allows to write
\bea
&&\hat \Pi^{5,0}_{\mu_1, \mu_2}(p)=\hat\Pi^{5,0,a} _{\mu_1, \mu_2}(p)+\hat\Pi^{5,0,b}_{\mu_1, \mu_2}(p)\nn\\
&&
\hat \Pi^{5,0}_{\mu_1, \mu_2,\mu_3}(p_1,p_2)=\hat\Pi^{5,0,a} _{\mu_1, \mu_2,\mu_3}(p_1,p_2)+\hat\Pi^{5,0,b}_{\mu_1, \mu_2,\mu_3}(p_1,p_2)
\eea
where the first term is obtained replacing all propagators
with ${\chi(k)\over-i \not k}$
and the second term is a rest, containing at least
an $r (k)$ term. More explicitly we have in $d=2$
\be
\hat \Pi^{5,0,a}_{\mu, \nu}=\int {dk\over (2\pi)^2} \Tr {\chi(k)\over-i \not k}\g_\m \g_5{\chi(k+p)\over-i \not k-i \not p}\g_\n
\ee
and in $d=4$
\bea
&&\hat \Pi^{5,0,a}_{\mu, \r,\s}(p_1,p_2)=i
\int {dk \over (2\pi)^4}\\
&&{\rm Tr} {\chi(k)\over \not k }\g_\m \g_5 {\chi(k+p)\over \not k+
\not p}\g_\n{\chi(k+p^2)\over \not k+\not p^2}\g_\s+[(p_1,\n)\xrightarrow  (p_2,\s)]\nn
\eea
We use now two crucial facts
\begin{enumerate}
\item $\hat \Pi^{5,0,a}_{\mu, \nu}(p)$
is not continuous in $p$ while $\hat \Pi^{5,0,b}_{\mu, \nu}(p)$
 is continuous; $\hat \Pi^{5,0,a}_{\mu, \r,\s}(p_1,p_2)$
is continuous in $p_1,p_2$ while $\hat \Pi^{5,0,b}_{\mu, \r,\s}(p_1,p_2)$
has
continuous derivatives.
\item The WI for the current implies that
$\sum_\n \s_\n\hat \Pi^{5,0}_{\mu, \nu}=0$
and $\sum_\r \s_\r\hat \Pi^{5,0}_{\mu, \r,\s}=\sum_\s \s_\s\hat \Pi^{5,0}_{\mu, \r,\s}=0$
\end{enumerate}
The computation of $\hat \Pi^{5,0,a}_{\mu, \nu}$ and $\hat\Pi^{5,0,a}_{\mu, \r,\s}$ is done using the
identity
\be
{\chi(k)\over \not k}\not p{\chi(k+p)\over \not k+\not p}=
[{\chi(k)\over \not k}-{\chi(k+p)\over \not k+\not p}]+
{\chi(k)\over \not k}C(k,p){\chi(k+p)\over \not k+\not p}
\ee
with
\be
C(k,p)=\not k (\chi^{-1}(k)-1)-(\not k+\not p) (\chi^{-1}(k+p)-1)
\ee
The momentum cut-off violates gauge invariance, and the correction
is given by the second term in the r.h.s. of (32).

Let us start considering $\hat \Pi^{5,0,a}_{\mu, \nu}$ in $d=2$; 
setting $\hat j_0=\hat \r_++\hat \r_-$, $\hat j_1=i(\hat \r_+-\hat \r_-)$,
$\hat j_0^5=\hat \r_+-\hat \r_-$, $\hat j^5_1=i(\hat \r_++\hat \r_-)$
so that $\hat\r_\pm$ the Fourier transform of $\psi^+_{\pm,x}
\psi^-_{\pm,x}$ and $\hat j_\m=i\e_{\m,\n}\hat j_\n$ ($\e_{0,1}=-\e_{1,0}=1$),
one
has to consider terms of the form
\be
<\hat \r_ {\o,p} \hat \r_ {\o,p} >= \int {dk\over (2\pi)^2}{\chi(k)\over D_\o(k)}{\chi(k+p)\over D_\o(k+p)}
\ee
with $D_\o(k)=-i k_0+\o k_1$ and $\o=\pm$. By using (31) and the
fact that
$\int {dk\over (2\pi)^2} 
 [{\chi(k)\over \not k}-{\chi(k+p)\over \not k+\not p}]=0$
we get
\bea
&&D_\o(p)<\hat \r_ {\o,p} \hat \r_ {\o,p} >=\int {dk\over (2\pi)^2}{1\over D_\o(k)}{1\over D_\o(k+p)}\nn\\
&&[D_\o(p) \chi(k)(1-\chi(k+p))-D_\o(k)(\chi(k)-\chi(k+p))]
\eea
By symmetry \be \int {dk\over (2\pi)^2}{1\over D_\o(k)^2}
\chi(k)(1-\chi(k))=0\ee and looking to the second term we write
\bea
&&-\int {dk\over (2\pi)^2|k|}{-i k_0 -\o k\over |k|^2}(p_0 k_0+p_1 k_1)\partial\chi=-D_{-\o}(p)\times\\
&&\int {dk\over (2\pi)^2}{k_0^2\over |k|^3}\partial\chi=-D_{-\o}(p){\pi\over (2\pi)^2}\int  d\r \partial\chi={D_{-\o}(p)\over 4\pi}\nn
\eea
In conclusion $D_\o(p)<\hat \r_ {\o,p} \hat \r_ {\o,p} >=
{D_{-\o}(p)\over 4\pi}$ , up to $O(a p^2)$, 
$p_\m \hat\Pi^{5,0,a}_{\mu, 0}=p_0 <(\hat\r_+ -\hat\r_-) 
(\hat\r_++\hat\r_-)>+i p_1
<(\hat\r_+ +\hat\r_-)( \hat\r_++\hat\r_-)>=i D_+ <\hat\r_+ \hat\r_+>-i D_- <\hat\r_- \hat\r_->=i(D_--D_+)=-2 i {p_1\over 2\pi}$ and
 $p_\m \Pi^{5,0,a}_{\mu, 1}=p_0 <(\hat\r_+ -\hat\r_-) i(\hat\r_+-\hat\r_-)>+p_1
<(\hat\r_+ +\hat\r_-)i(\hat \r_+-\hat\r_-)>=-D_+ <\hat\r_+ \hat\r_+>- D_- <\hat\r_- \hat\r_->=-(D_- + D_+)=2 i {p_0\over 2\pi}$ so that 
\be p_\m \hat\Pi^{5,0,a}_{\mu, \nu}=-i {\e_{\mu, \nu}\over 2\pi}p_\m\ee
Moreover 
$p_\n \hat\Pi^{5,0,a}_{0, \n}=p_0 <(\hat\r_+ -\hat\r_-)(
 \hat\r_++\hat\r_-)>+i p_1
<(\hat\r_+ +\hat\r_-)( \r_+-\r_-)>=i D_+ <\hat\r_+ \hat\r_+>-i D_- <\hat\r_- \hat\r_->=i (D_--D+)=-2 i {p_1\over 2\pi}$;
$p_\n \hat\Pi^{5,0,a}_{1, \n}=i p_0 <\hat\r_+ +\hat\r_-:\hat \r_+ +\hat\r_->- p_1
<\hat\r_+ +\hat\r_-: \hat\r_+-\hat\r_->=-D_+ <\hat\r_+ \hat\r_+>- D_- <\hat\r_- \hat\r_->=-(D_-+D+)=2 i {p_0\over 2\pi}$ so that
\be
p_\n \hat\Pi^{5,0,a}_{\mu, \nu}=-i{\e_{\nu, \mu}\over 2\pi}p_\n+O(a p^2)
\ee
Note that the current and the axial current are not conserved if we consider only the continuum part of the propagator (27) with momentum cut-off.
On the other hand, from the lattice WI we get
\be 
i \s_\n \hat\Pi^{5,0}_{\mu, \nu}={\e_{\nu, \mu}\over 2\pi}p_\n+i \s_\n \Pi^{5,0,b}_{\mu, \nu}(p)+
O(a p^2)\label{ipp}
\ee
We use the continuity of $\hat\Pi^{5,0,b}_{\mu, \nu}(p)$ to conclude from \pref{ipp} that
$i\hat\Pi^{5,0,b}_{\mu, \nu}(0)=-{\e_{\nu, \mu}\over 2\pi}$; therefore
so that $
i\s_\m \hat\Pi^{5,0}_{\mu, \nu}$
\be=i\s_\m \hat\Pi^{5,0,a}_{\mu, \nu}+i\s_\m \hat\Pi^{5,0,a}_{\mu, \nu}={1\over 2\pi}
(\e_{\mu,\nu}- \e_{\nu, \mu})+O(a p^2)
\ee
that is $i\s_\m \hat\Pi^{5,0}_{\mu, \nu}={\e_{\nu, \mu}\over \pi}+O(a p^2)
$.

We can follow a similar strategy also in $d=4$. By using
(31) one obtains (see 3.6 of \cite{54})
\bea 
&&-i (p_{1,\mu} + p_{2,\mu}) \hat \Pi^{5,0,a}_{\mu,\nu,\sigma}
 = {1\over 6\pi^{2}} p_{1,\alpha} p_{2,\beta} \varepsilon_{\alpha\beta\nu\sigma}+O(a \bar p^2)\nn	\\
&&
-i p_{1,\nu}  \hat \Pi^{5,0,a}_{\mu,\nu,\sigma}={1\over 6\pi^{2}} p_{1,\alpha} p_{2,\beta} \varepsilon_{\alpha\beta\mu\sigma}+O(a \bar p^1)
\eea
with $\bar p=\max (|p_1|,|p_2|)$.
The WI for the current $p_{1,\nu}  \hat \Pi^{5}_{\mu,\nu,\sigma}=0$, that is 
\be
{1\over 6\pi^{2}} p_{1,\alpha} p_{2,\beta} \varepsilon_{\alpha\beta\mu\sigma}+p_{1,\nu}  \hat \Pi^{5,0,b}_{\mu,\nu,\sigma}(p_1,p_2)+
O(a \bar p^2)=0
\ee
Using the continuous differentiability of $\hat \Pi^{5,0,b}_{\mu,\nu,\sigma}(p_1,p_2)$ we can expand in series and equating the coefficients
\begin{equation}\label{id.1}
\hat \Pi^{5,0,b}_{\mu,\nu,\sigma}(0, 0) = 0\;,\qquad \frac{\partial \hat \Pi^{5,0,v}_{\mu,\nu,\sigma}}{\partial p_{2,\beta}} (0, 0) = - \frac{1}{6\pi^{2}}\varepsilon_{\nu\beta\mu\sigma}
\end{equation}
Similarly, $\frac{\partial \widetilde H^{5,0,b}_{\mu,\nu,\sigma}}{\partial p_{1,\beta}}(0,0) = -\frac{1}{6\pi^{2}}\varepsilon_{\sigma\beta\mu\nu}$
so that expanding $p_\m=p^1_\m+p^2_\m$
\begin{eqnarray} &&\label{WIfinfin3}
-i p_\m\hat\Pi^{5,0}_{\mu,\nu,\sigma}=
{1\over 6\pi^{2}} (p_{1,\alpha} p_{2,\beta} \varepsilon_{\alpha\beta\nu\sigma}-p_{1,\alpha} p_{2,\m}\varepsilon_{\s\alpha\m\n}\\
&&-p_{1,\m} p_{2,\b}\varepsilon_{\nu\b\m\sigma})+O(a \bar p^2)={1\over 2\pi^{2}} p_{1,\alpha} p_{2,\beta} \varepsilon_{\alpha\beta\nu\sigma}+O(a \bar p^2)\nn
\end{eqnarray}
The value of the anomalies with finite lattice is therefore the same
as in the continuum case up to subdominat terms. The same property is true for
the transport coefficients; for instance the optical conductivity
in graphene is the same on the lattice or in the
continuum description \cite{53}.

\subsection{Renormalization Group analysis}

Let us consider now the interacting case. In order to
study the properties of anomalies, as well as other other
physical quantities, we need to express the correlations
in terms of expansions, which will be in general not
power series. Such series are obtained by a Renormalization
group analysis and are expected to be convergent at
most only in certain regions of the parameters. A case in
which a convergent expansion can be found is under the
condition that $e^2/M^2 a^2$ is small,  see \cite{55},\cite{56}, 
\cite{57}.

We start considering $W(J , J^5,0)$, $r=1$,  (6) which can be
rewritten more compactly as
$\int P(dA)
P(d\psi) e^{V(\psi, A+J,J^5)}$.
One can integrate out the boson field reducing to a purely
fermionic theory
\be
\int P(dA)
P(d\psi) e^{V(\psi, A+J,J^5)}=\int P(d\psi) e^{\bar V(\psi, J,J^5)}
\ee
with 
\be \bar V=\sum_{l,m}  a^{4 (l+m)} 
\sum_{\underline x,\underline y}   H_{l,m} (\underline x,\underline y) 
\prod_{i=1}^l \psi^{\e_i} _{x_i}\prod_{j=1}^m J^{a_i} _{y_i}
\label{aaa} \ee 
where $\e=\pm$, $a=0,5$ and $J^0=J$. The kernels $H_{l,m} (\underline x,\underline y) $
are given by
{\it Truncated expectations}  $\EE^T_A( e^{i\e_1 e   a 
A_{\m_1} };...)$. 
One can use a suitable representation of truncated expectations in terms of sum over trees, see \cite{58}
\be
\EE^T_A( e^{i\e_1 e   a 
A_{\m_1} };...; e^{i\e_n e   a 
A_{\m_n}})=\sum_{T }\prod_{i,j\in T} V_{i,j}
\int dp_T(s) e^{-V(s)}
\ee 
with 
$T$ are connected tree graphs on $\{1,2,\ldots,n\}$, the product $\prod_{\{i,j\}\in T}$ runs over the edges of the tree graph $T$,
$\int dp_T(s)=1$, $V(s)$ convex combination of $V(Y)=\sum_{i,j\in Y}V_{i,j}$, $Y$ subsets of $X$,
$V_{i,j}= \EE( a A_{\m_i} (x_i)
a A_{\m_j}(x_j ))$.By using that  $V(Y)\ge 0$,
$V(s)\ge 0$ and $\sum_T\le C^n n!$  we get  for $e/(M a)$ small enough
\be||H_{l,m}||
\le C^{l+m} 2^{(4-3 l/2-m)N}  \ee
with $||H_{l,m}||={1\over L^4} a^{4(l+m)}\sum_{\underline x,\underline y}
|H_{l,m}(\underline x,\underline y)|$. Note that by using (47) one is avoiding to
expand in $e$, what is not suitable for a non-perturbative
analysis.
We are then reduced to a purely fermionic theory. The
fermions are however massless, so that a convergent expansion
can be obtained only by a multiscale Renormalization
Group analysis.

We introduce a partition of the unity
$\sum_{h=-\io}^{N+1}  f_h(k)=
1$ with $f_h(k)$, $h\le N$,  a smooth compact support cut-off function non vanishing for $2^{h-1}\le |k|_T\le 2^{h+1}$, with $|k|_T$ the periodic norm in $[-\pi/a,\pi/a]^4$ and $2^N=1/(10 a)$.  
We can write the propagator as sum of
propagators living at decreasing momentum scales. If $\hat g^h(k)=f_h(k)\hat g(k)$
\be
\hat g(k)=\sum_{h=-\io}^{N+1} \hat g^h(k) \quad\quad \hat g^{\le h}(k)=\sum_{i=-\io}^h \hat g^i(k) 
\ee
\insertplot{260}{69}
{\ins{10pt}{18pt}{$\EE^T_{N+1}$}
\ins{50pt}{18pt}{$V^{N+1}$}
\ins{60pt}{30pt}{$+$}
\ins{90pt}{15pt}{$\EE^T_{N+1}$}
\ins{120pt}{0pt}{$V^{(N+1)}$}
\ins{120pt}{60pt}{$V^{(N+1)}$}
\ins{130pt}{30pt}{$+$}
\ins{170pt}{15pt}{$\EE^T_{N+1}$}
\ins{210pt}{0pt}{$V^{(N+1)}$}
\ins{210pt}{60pt}{$V^{(N+1)}$}
\ins{210pt}{30pt}{$V^{(N+1)}$}
\ins{250pt}{30pt}{$+$}
\ins{260pt}{25pt}{$...$}
}
{figjsp44abb}{ 
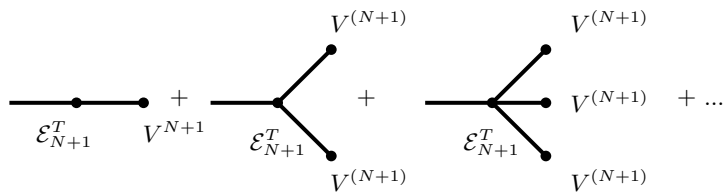
Graphical representation of $V^N$; the first term represents $\EE^T_{N+1}(V^{(N+1)})$, the second 
${1\over 2}\EE^T_{N+1}(V^{(N+1)};V^{(N+1)})$ and so on.
}{0}

By using the additivity property of
Grassmann Gaussian integrals, see e.g. \cite{59}, we can write, calling $\bar V\equiv V^{N+1}$
\be
\int P(d\psi) e^{V^{N+1}(\psi, J,J^5)}=
\int P(d\psi^{(\le N)}) e^{V^N( \psi^{(\le N)}, J,J^5))}\label{llla}
\ee
with $e^{V^N( \psi^{(\le N)}, J,J^5))}=\int P(d\psi^{N+1}
) e^{V^{N+1}(\psi, J,J^5)}$. Again
$V^N$ is expressed by the sum of fermionic truncated expectations
$V^N=\sum_{n=0}^\io {1\over n|}\EE^T_{N+1}(V^{N+1}, ...,V^{N+1})
$, 
and this expansion
can be conveniently represented in Fig.1. The effective
potential $V^N$ is sum over monomlals with $l$ $\psi$  fields
and $m$ J-fields.
One separates in $V^N$ the part containing marginal or relevant
terms from the rest and the procedure can be iterated;
after the integration of the fields $\psi^N,\psi^{N-1},..,\psi^{h+1}$
one obtains
\be
\int P(d\psi) e^{V^{N+1}(\psi, J,J^5)}=
\int P(d\psi^{(\le h)}) e^{V^h(\sqrt{Z_h} \psi^{(\le N)}, J,J^5))}\label{llla}
\ee
where $P(d\psi^{(\le h)})$ has propagator ${\hat g^h(k)\over Z_h}$, with $Z_h$ the wave function renormalization and
the effective potential $V^h$ is given by 
$V^h(\sqrt{Z_h} \psi^{(\le h)}, J,J^5)=$
\bea
&&a^4\sum_x ( 2^h \n_h \bar\psi_x\psi_x+Z_h^J J_\m \bar\psi_x\g_\m \psi_x+
Z_h^5 J^5_\m \bar\psi_x\g_5\g_\m \psi_x)\nn\\
&&+\sum^*_{l,m}  a^ {4(l+m)} \sum_{\underline x,\underline y} 
H_{l,m}^h(\underline x, \underline y) \prod_{i=1}^l \partial^{q_i}\psi^{\e_i} _{x_i}\prod_{j=1}^m  J_{y_i}\label{aaa22}
\eea
and $\sum^*$ is over the terms with negative dimension $D<0$ where $D=4-3l/2-\sum_i q_i$; the first line in \pref{aaa22}
contains the marginal or relevant terms and the second
the irrelevant terms. 

\insertplot{300}{160}
{\ins{60pt}{90pt}{$v_0$}\ins{120pt}{100pt}{$v$}
\ins{100pt}{90pt}{$v'$}
\ins{120pt}{-5pt}{$h_v$}
\ins{235pt}{-5pt}{$N$}
\ins{255pt}{-5pt}{$N+1$}
}
{treelut2}{\label{n11} A Gallavotti-Nicolo' tree 
}{0}

The effective potential
$V^h$ is expressed as sum of truncated expectations of
$V^{h+1}$
 which also are sum of truncated expectations of
$V^{h+2}$ and so on; therefore iterating the graphical representation
in Fig. 1 one obtains that the kernels
$H_{l,m}^h$
can be written
in terms of Gallavotti-Nicolo' trees \cite{26} (see e.g. \cite{59}), see Fig.2.

The tree expansion allows to write the kernels in terms
of the running running coupling constants (rcc) 
$\n_j, Z^J_j,Z^5_j$
with $h\le j \le N$ and in $e$. Such series are indeed convergent,
as follows by 
determinant bounds for fermions and by a convenient
representation for fermionic truncated expectations \cite{58}
(see also e.g.  \cite{59})
$
\EE^T_{h}(\tilde\psi^{(h)}(P_1);
...;\tilde\psi^{(h)}(P_s))=$
\be
\sum_{T}\prod_{l\in T} g^{(h)}(x_l,y_l)
\int dP_{T}(t) \det
G^{h,T}(t)
\ee
where $\psi(P)$ are monomials in the fields $\psi$,
$T$ is 
$G^{h,T}(t)$ is a $(n-s+1)\times (n-s+1)$ matrix
which is bonded by Gram inequality, $n$ the number of fields. Again avoiding
the use of Feynman graphs eliminate factorials in the bounds in the bounds preventing convergence.

The running coupling constants remain small at any
iteration step, provided that the counterterm $\n$ is properly
chosen, and $\n_h\to 0$, $Z_h\to Z,Z^J_h\to Z^J,Z^5_h\to Z^5$  
as $h\to-\io$; such limits are expressed
by non trivial series expansions depending on all lattice
details. The propagator can be written as
\be
\hat g^h(k)={f^h(k)\over -i \not k}+\hat r^h(k)\ee with 
$\hat r^h(k)/\hat g^h(k)=O(2^{h-N})$: the single scale propagator is 
equal to the continuum propagator up to correction.

We can write 
$H_{n,m}^h=H_{a,n,m}^h+H_{b,n,m}^h$ where $H_{a,n,m}^h$ 
includes only contributions from $Z_h^J, Z^5_h$ terms and from the leading part of the propagator ${f^h(k)\over -i \not k}$
and 
$H_{b,n,m}^h$ 
are the other terms. The outcome of the analysis is the following \cite{54}
\vskip.3cm
\noindent
{\it For ${e^2\over 
(M a)^2}\le e_0$, where $e_0$ is constant and for a suitable $\n$
\be
||H_{a, l,m}^h||\le C^{l+m}  2^{D h} \quad\quad 
||H_{b, l,m}^h||\le C^{l+m} 2^{D h} 2^{\th (h-N)}  \label{speror}
\ee
with $\th>0$, $D=4-3l/2-m$; moreover $|Z-1|, |Z^J-1|, |Z^5-1|\le C e^2$
}

The above result ensures that the expansions can
be used to get non-perturbative information if ${e^2\over 
(M a)^2}$
is
smaller than some numerical constant $e_0$, independent
from $L$ ; this is the typical condition appearing in nonrenormalizable
theories like the Fermi theory of weak interactions
in $d=4$ (whose convergence is a corollary
of the above result).
The explicit value of $e_0$ can be deduced
collecting all the numerical constants in the proof
of convergence, but no attempt to optimize it has been
given. The series so obtained are not power series in the
coupling $e$; the propagator at each scale $g^h$ have a non
trivial dependence on the coupling $e$, and the wave function
or vertex renormalization have a dependence on
the scale. Convergence ensures that physical quantities
can be obtained by lowest order in this expansion with
a bound on the remainder. Note the improvement in the
estimate for the $H_{b, l,m}^h$
 terms, which follows from the fact
that only irrelevant sense in the RG sense contribute. Of
course, the above expansion is not the only possible one
and others can be proposed; in particular one could decompose
the $A_\m$ field in scales instead of integrating it out,
and hopefully convergence up to higher cut-off is found,
see below.

\subsection{Emergent symmetries and non-renormalization
of the anomaly}

As a consequence of the above analysis one obtains
\bea
&&\hat S(k)={1\over Z \not k}(1+r_1)\nn\\
&&G_\m(k,p)={Z Z^J\over Z^2} \hat g(k)\g_\m\hat g(k+p) (1+r_2)\nn\\
\label{lal}\\
&&G^5_\m(k,p)={Z Z^5 \over Z^2} \hat g(k)\g_\m\g_5\hat g(k+p) (1+r_3)\nn
\eea
where $|r_1|\le C (a |k|)^\th$, $|r_2|,|r_3|\le C (|k|^\th+|k+p|^\th)$, see Fig.3. The first term comes from the 
$H_{a, l,m}^h$
and the second from the
$H_{b, l,m}^h$; the momentum dependence
is due to the factor $2^{\th (h-N)}$
 in the bound. The
terms $r_i$ are small at energies far from the cut-off,
that is Lorentz symmetry emerges at low energies. The
fact that the dominant term is identical to the free correlation
with renormalized parameters comes from the
fact that all the non-irrelevant terms are quadratic in the
fermions.

\insertplot{230}{70}
{\ins{50pt}{40pt}{$=$}
\ins{130pt}{40pt}{$+$}}
{figjsp44e}
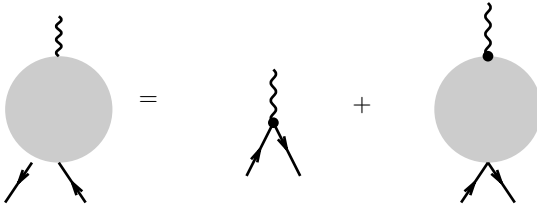
{\label{h2} Graphical representation of the seconf of (55); the dot represent $Z^J$, the first term in the r.h.s. is the renormalized vertex with two renormalized propagators, and the second is the contributions of higher order terms with at least an irrelevant term} {0}

The renormalization $Z,Z^J$ depend on all the lattice
and interaction details; however combining the Ward
identities (19) we get the following identity
\be
{Z^J\over Z}=1
\ee
implying the non-renormalization of the electric charge.
This is a peculiarity of the lattice regularization; with momentum
regularization this would be not true.

We also choose $\ZZ^5$ so that $Z^J/Z^5=1$, as required by the
definition of the axial current. This implies that
\be\frac{Z^{5} Z^{J} Z^{J}}{Z^3 }=1\ee

Let us consider now the three point correlation for the
chiral current, which can be written
\be
\Pi^5_{\m\n\s}=\sum_{h=-\io}^N \hat H_{a,0,3}^h+\sum_{h=-\io}^N \hat H_{b,0,3}^h
\ee
where again the first term in the l.h.s. is given  by
marginal terms and the dominant part of the propagators, hence is given by sum of triangle graphs
\bea
&&\hat H_{a,0,3}^h(p_1,p_2)=i \sum^*_{h_1,h_2,h_3}
{Z_{h_1}^5\over Z_{h_1}}{Z_{h_2}^J\over Z_{h_2}}{Z_{h_3}^J\over Z_{h_3}}
\int {dk \over (2\pi)^4}\\
&&{\rm Tr} {f_{h_1}(k)\over \not k }\g_\m \g_5 {f_{h_2}\over \not k+
\not p}\g_\n{f_{h_3}(k+p^2)\over \not k+\not p^2}\g_\s+[(p_1,\n)\xrightarrow  (p_2,\s)]\nn
\eea
where $\sum^*_{h_1,h_2,h_3}$ means that
at least one among $h_1.h_2,h_3$
is equal to $h$. In contrast, the second term is given a complicate
series,contaning at least an irrelevant term, see
Fig.3. $\hat H_{b,0,3}^h$
is not small as funtion of momenta;
however is more regular than $\hat H_{a,0,3}^h$. By
(55) we see that $\Pi^5_{\m\n\s}$ 
is continuous as is bounded by $\sum_{h=-\io}^N 2^h
$; moreover each derivatives
produces an extra $2^{-h}$ 
so that $\sum_{h=-\io}^N \partial \hat H_{b,0,3}^h$ 
is bounded by $\sum_{h=-\io}^N 2^{\th(h-N)}$
hence $\hat H_{b,0,3}^h$
has continuous derivatives. In addition
$Z_h\to Z,Z^J_h\to Z^J,Z^5_h\to Z^5$
exponentially fast
; performing the
sum over $h_1,h_2,h_3$
in the first term reconstruct 
$\Pi^{5,a.0}_{\m\n\s}
$ in (30) so that, see Fig. 3
\be
\hat\Pi^{5}_{\mu,\nu,\sigma}(p_1,p_2)= \frac{Z^{5}Z^{J} Z^{J}}{Z^3 } \hat\Pi^{5,0,a}_{\mu,\nu,\sigma}(p_1,p_2)
+\hat\Pi^{5,0,b}_{\mu,\nu,\sigma}(p_1,p_2)\label{azz}\ee
We can proceed now exactly as we did in the on interacting case in (41)-(43): even if $\hat\Pi^{5,0,b}_{\mu,\nu,\sigma}$
 is an all order expansion,it is still differentiable in the momenta so that we
can expand up to first order in the momenta.
Using (58)
and the fact that the first derivatives of $\hat\Pi^{5,0,b}_{\mu,\nu,\sigma}(p_1,p_2)$ are fixed by the WI
for the current $\s_\n\hat\Pi^{5}_{\mu,\nu,\sigma}=0$
as in (43), and using (41)
we get the following result,
see \cite{60},  \cite{62}.

\insertplot{230}{70}
{\ins{50pt}{40pt}{$=$}
\ins{130pt}{40pt}{$+$}
}
{figjsp44a}
{\label{h2} Graphical representation of \pref{azz}.
} {0}

\vskip.3cm
\noindent
{\it If $e^2\le  
\e_0 (M a)^2$ and for a suitable $\n,\ZZ^5$, if $|p|=max(|p_1|,|p_2|)$, $\th>0$
\be
-i p_\m \Pi^5_{\m,\r,\s}={1\over 2\pi^2}
\e_{\a, \b,\r,\s} p^1_\a  p^2_\b+O(a^\th |p|^{2+\th})\ee}

The anomaly is therfore non-renormalized with a finite lattice and in presence of interaction with a massive $U(1)$ gauge field.

\section{Universality of transport in condensed matter}

\subsection{Weyl semimetals}


Weyl semimetals are condensed matter systems with an
emerging QED$_4$ description [17] subject to an intense experimental
study \cite{50}. A basic model for a Weyl semimetal can be expressed
in terms of fermions hopping on a lattice with
suitable complex weights and a current-current interaction;
the generating function is

\be
e^{W(J, J^5, \phi)}=\int P(d\psi) e^{V(\psi)+A(J)+\n N+(\psi, \phi)+(j^5_\m(J), J^5)}
\ee
where $\psi^\pm_x=(\psi^\pm_{x,a}, \psi^\pm_{x,b})$,
$x=(x_0, \vec x)$ with $x_0$ the imaginary time $x_0\in (0, \b)$ (antiperiodic boundary conditions) and $\vec x
\in \L$ a square lattice with step $1$, that is $\vec x\in [0,L]^3
\cap \ZZZ$.
$P(d\psi)= \DD \psi e^{S_0(\psi)}$ has propagator $\hat g(k)=$
\be
\begin{pmatrix} -ik_0 +t_1 \sin k_1-it_2\sin k_2
 & \h-\cos k_1-\cos k_2-\cos k_3 \\ \h-\cos k_1-\cos k_2-\cos k_3 & -ik_0-(t_1 \sin k_1-it_2\sin k_2)\end{pmatrix}^{-1}
\ee
with $\h=2+\z$
and $t_1,t_2>0$; moreover $V$ is a density-density interaction, if  ${1\over \b}\int dx_0  \sum_{\vec x\in \L}=\int dx$
\be
V=\l\int dx dy\sum_{i,j}
 v(\vec x,\vec y)\d(x_0-y_0) \psi^+_{x,i}\psi^-_{x,i}
\psi^+_{y,i}\psi^-_{y,i}
\ee
with $v(\vec x, \vec y)$ a short range interaction.
$A(J)$ is the source term for the current, obtained from the fermionic action $S_0(\psi)$ 
replacing $\psi^+_x \psi_{x+e_i}$ with 
$\psi^+_x (e^{J^i_x}-1) \psi_{x+e_i}$ and replacing $\partial_0$ with $A_0$. This choice ensures the validity of Ward Identities for the current.
Finally
$N=\int dx (\psi^+_{x,1}\psi^-_{x,1}-\psi^+_{x,2}\psi^-_{x,2})$
and $\n$ is a counterterm necessary to fix the chemical potential.

For 
$-1<\zeta<1$ the denominator of $\hat g(k)$ is
vanishing in correspondence only of two points  (Weyl nodes, or Fermi points) 
$k=p_F^\pm$, with $p_F^\pm=(0,0,\pm\arccos\zeta)$. The relative distance between the two nodes vanishes like $\sqrt{1-|\zeta|}$ as 
$|\zeta|\to 1^-$.
In the vicinity of the Weyl nodes, $k\simeq p_F^\omega$, the propagator can be approximated by, if $v_F=\sqrt{1-\zeta^2}$
\begin{equation}\label{eq:linear}
 (-i k_0 v_1^0\sigma_1 k_1+v_2^0 \sigma_2 k_2 +\omega v_F\sigma_3 (k_3-p_{F,3}^\omega)
+ O(|k-p_F^\o|^2)^{-1}\;,
\end{equation}
with $v_1^0=t_1, v_2^0=t_2$.
One has, 
depending if $\o=\pm$, a propagator approximately given by
the propagator of
$L$ or $R$ massless Dirac fermions with an anisotropic velocity
and up to subleading corrections.
Finally 
one introduces a lattice current for the quasi-particle flow between the Weyl nodes
\begin{equation}\label{eq:J5def}
\hat j^{5}_{\mu,p} = \frac{\ZZ^{5}_{\mu}}{L^{3}} \sum_{k} \hat 
\psi^{+}_{k+p} w^{5}_{\mu}(k,p) \hat \psi^{-}_{k}
\end{equation}
with
$w^{5}_{\mu}(k,p) $ a suitable kernel
ensuring that $\hat J^{5}_{\mu,p}$ reduces to the chiral relativistic current at small momenta
and renormalization $\ZZ_5^\m$  so that
the charge is the same as the one carried by the current. Note in particular $\hat j^{5}_{0,p} $
represents the difference of densities of electrons around the Weyl points.
Finally
$ j^{5}_{\mu}(J)$ is defined so that
$ j^{5}_{\mu}(0)$ is equal to (67) and invariance under the transformation
$\psi^\pm\to \psi^\pm e^{\pm i e \a_x}$, $J_\m\to J_\m+d_\m \a$
is valid.

$\Pi^5_{\m\n\r}$ is the derivative with respect to $J^5_\m, J_\n,J_\r$
and  represents the quadratic response
of the expectation of the chiral 4-current $j^5_\m$ with
respect to the external fields. In the non-interacting case 
one can proceed as in \S 2.4 obtaining that the time variation 
of the difference of densities between Weyl nodes is 
$\frac{1}{2 \pi^{2}} \int dx\, E(t) \cdot B(t)
$
up to an error term, collecting contributions involving a
higher number of derivatives on the vector potential, 
which is subdominant for a vector potential slowly varying in
space. The quasi-particle flow is therefore proportional
to $E(t) \cdot B(t)$, with coefficient given by the value of the chiral
anomaly.
It is important to predict if this value
is or not renormalized by interactions, in view of possible
comparison with experiments.

A multiscale analysis can be performed \cite{54}, \cite{61}.
The fermionic propagator is written as $\hat g(k)=
\hat g^{\ge \bar h}(k)+\hat g^{\le \bar h}(k)$, where 
the scale $\bar h$ is fixed by the condition that $v_F (k_3-p_F^\pm)=O( |k_3-p^\pm_F|^2)$ in the support of $f_{\bar h}$ . In the first region $k_3$ has a quadratic scaling and in the second the support decouples and 
$\hat g^{\le \bar h}(k)$ is disconnected in two regions around
the 2 Fermi points with linear scaling: therefore we can write 
$\psi^{\le \bar h}=\sum_{\o=\pm 1} \psi^{\le \bar h}_\o$ and the RG analysis is in terms of Dirac fermions with anisotropic
velocities. 
The  value of the Weyl ponts is also modified in
presence of an interaction and one has to fix the chemical potential
properly choosing the counterterm $\n$ so that their
value is the same in the free and interacting case.

Even if the emerging theory is still described by massless
Dirac fermions, the deviation from a relativistic theory
is surely much more drastic as the one in the previous section.
First the velocity is anisotropic, and one velocity is much
smaller than the others; moreover the two Dirac cones
are connected in the dispersion relation and the interaction
only involve the densities. The result of the analysis
is nevertheless that the chiral current correlation can be
written in a form similar as before, for $\l$ small uniformly
in the Fermi velocity
\begin{equation}\label{eq:tritetr_intro}
\hat\Pi^{5}_{\mu,\nu,\sigma}(p_1,p_2)= \frac{Z^{5}_{\mu} Z_{\nu} Z_{\sigma}}{Z^3 v_1v_2v_3} \Pi^{5,a,0}_{\mu,\nu,\sigma}(\lis p_1,\lis p_2)+\widetilde H^5_{\mu,\nu,\sigma}(p_1,p_2),
\end{equation}
where the first term is given by the triangle graph with momentum cut-off 
and velocities equal to $1$
but 
computed at 
$\lis p_1=(p_{1,0}, v_1 p_{1,1}, v_2p_{1,2}, v_3p_{1,3})$, 

$\lis p_2=(p_{2,0}, v_1 p_{2,1}, v_2p_{2,2}, v_3 p_{2,3})$; the second term is a series of terms,
which is again differentiable in the external momenta. 
Note that $v_1,v_2,v_3$ are non trivial functions of the interaction $\l$.
Ward Identities hold so that
\begin{equation}
Z_{\mu} = v_{\mu} Z\;,\label{asympt.2_intro}
\end{equation}
with $v_0:=1$, and we impose $Z_\m^5=Z_\m$ by a suitable choice of $\ZZ_\m^5$.
We use the WI to 
compute the first derivative of the correction term, so that \cite{54}, \cite{61} the following result is obtained.
\vskip.2cm
\noindent
{\it For $|\l|\le \l_0$, with $\l_0$ a constant indendent
on $v_F$,  for a suitable $\n.\ZZ^5$ we get
\be
p_\m \Pi^5_{\m,\r,\s}=-{1\over 2\pi^2}
\e_{\a, \b,\r,\s} p^1_\a  p^2_\b+O( |p|^{2+\th})\ee
}

Despite the strong deviation from a relativistic theory due to lattice
and non-relativistic terms.
the quadratic response of the quasi-particle flow between
the Weyl nodes
, simulating the chiral anomaly, is still
perfectly non-renormalized with short range interactions. This is in contrast with other
quadratic responses in Weyl semimetals which show indeed
interaction dependent corrections \cite{51}. Experiments in Weyl semimetals are still not sufficiently precise to verify the non-renormalization.

\subsection{Graphene}
In Graphene the conduction electrons are described by
fermions hopping in a honeycomb lattice.
If $\L$ is a periodic triangular lattice with basis
${\vec a_1}={1\over 2}(3,\sqrt{3})$, ${\vec a_2}={1\over
2}(3,-\sqrt{3})$, the propagator is given by
\be
\hat g(k)=
\begin{pmatrix}
i k_0 & -v^*(\vec k) \cr -v(\vec k) &
ik_0
\end{pmatrix}^{-1}
\ee
where $k=(k_0,\vec k)$ and $k_0={2\pi\over \b}(n_0+{1\over 2})$ and $\vec k={n_1\over L}{\vec b_1}+ {n_2\over L}{\vec
b_2}$, where $\vec b_1= {2\pi\over
3}(1,\sqrt{3})$, ${\vec b_2}={2\pi\over 3}(1,-\sqrt{3})$. Finally 
$
v(\vec
k)=t \sum_{i=1}^3e^{i\vec k(\vec\d_i-\vec\d_1)}=t(1+2 e^{-i 3/2
k_1}\cos{\sqrt{3}\over 2}k_2)$,$t$ the hopping parameter.
If we take $\b,L\to\io$, the limiting propagator $\hat g(k)$ becomes
singular at $k_0=0$ and $\vec k=\vec p_F^{\pm}$,
where
$\vec p_{F}^{\ \pm}=({2\pi\over 3},\pm{2\pi\over 3\sqrt{3}})$. 

The
asymptotic behavior of $v(\vec k)$ close to the Fermi points is given by
$v(\vec p_F^\pm+\vec k')\simeq 3/2 t(i k_1'\pm k'_2)$. In particular,
if $\o=\pm$, the Fourier transform of the 2-point Schwinger function
close to the Fermi point $\vec p_F^\o$
can be rewritten in the form:
$\hat g(k_0,\vec p_F^\o+\vec k')=$
\be
\begin{pmatrix}
-i k_0 & -v_F^0(-ik_1'+\o k_2') +r_\o\cr -v_F^0
(ik_1'+\o k_2') +
r_\o^*& -ik_0
\end{pmatrix}^{-1}
\;,\label{1.8}\ee
where $v_F^{(0)}=3/2 t$
is the free Fermi velocity. Moreover,
$|r_\o(\vec k')|\le C \big|\vec k'|^2$, for small values of $\vec k'$
and for some positive constant $C$. 
Combining the two point function around the two Fermi points we get the propagatot
of Dirac particles in $d=2+1$ up to subdominant corrections.

If $\Pi_{i,j}$ is the current-currenti correlation with component $i,j$,
with $i$ and $j$ equal to $1$ or $2$,
the optical conductivity is given by, in the $U=0$ non interacting case,
in the $L,\b\to\io$ limit
\be
\s_i=\lim_{p_0\to 0} \lim_{p\to 0}  {1\over p_0}\Pi_{i,i}(p)={e^2\over h}{\pi\over 2}\label{opt}
\ee
Note
that $\Pi_{i,i}(p)$
is an even function $\Pi_{i,i}(p)=\Pi_{i,i}(-p)$ and $\Pi_{i,i}(0)=0$
; hence $\s_i$ is the derivative in $p=0$ and the fact
that is non vanishing is due to the fact that $\Pi_{i,i}(p)$ is
not differentiable. The derivation of \pref{opt} with a finite lattice can be done by direct computation
\cite{53} or proceeding as in \S 2.4 writing decomposing 
$\Pi_{i,i}(p)$ in a non differentiable and differentiable part and using WI.
The same value is found in Dirac approximation, see \cite{L}.

The optical conductivity $\s_i$ is
independent from the microscopic parameters in the non-interacting case, in particular from the hopping parameters $t$.
Experiments show a value very close, up to experimental
errors, to ${e^2\over h}{\pi\over 2}$ \cite{14} ; this requires an explanation
as interactions are rather strong and their presence is
known to modify other quantities,like the Fermi
velocity; which has been measured and show a strong
increase due to interactions \cite{67}.

Graphene with short
range interactions is described by the Hubbard model
on the honeycomb lattice. The generating function of the
correlation is
\be
\int P(d\psi) e^{V(\psi)+B(J,\psi)}
\ee
wit, if $\r_x$ is the density
\be V=\int dx dy
v(\vec x,\vec y)\d(x_0-y_0) \r_x \r_y\ee
and $B(J,\psi)=\int 
dx_0 \sum_x A_0 J_0+B_1$ with $B_1$ the source term for the current. Again exact WI holds and, after performing a similar RG analysis
%
the interacting two point functions is given by, $\o=\pm$ 
\be
\frac1{Z}
\begin{pmatrix}
-i k_0 & -v_F(-ik_1'+\o k_2')\cr -v_F
(ik_1'+\o k_2') & -ik_0\end{pmatrix}^{-1} \Big(1+ R(k')\Big)\;,
\label{1.9}\ee
with $k'=(k_0,\vec k')$, and with $Z$ and $v_F$ two real constants such that
\be Z=1+ a U^2+O(U^3)\;,\qquad\qquad v_F=\frac32 t+
b U+O(U^3)\label{1.10}\ee
and  $|R(k')|\le C |k'|^\theta$.
The effect of the interaction is to renormalize the Fermi velocity;
such renormalization would be absent in a relativistic model. Moreover
due to symmetries the velocity is shifted in an isotropic way. The scaling dimension is $D=3-l-m$ hence as in the previous case
the quartic terms in the fermions are irrelevant.

It is found that $\Pi_{i,i}(p)$ in the $L,\b\to\io$ limit
 is continuous and non differentiable; by using the WI, $i\not=0$ 
\be
\lim_{p\to 0}\lim_{p_0\to 0} \Pi_{ii}=0
\ee
and by continuity 
\be
\s_i=
\lim_{p_0\to 0} \lim_{p\to 0}{1\over p_0}(\Pi_{i,i}(p)-\Pi_{i,i}(0))\ee
As an outcome of the RG analysis
we can decompose $\hat \Pi_{i,i}(p)$, 
as
\be \hat\Pi_{lm}(p)= \frac{Z_l Z_m}{Z^2}
\media{\hat j_{p,l};\hat j_{-p,m}}_{0,v_F}+\hat
R_{lm}(p)\nn\ee
where $\media{\cdot}_{0,v_F}$ is the average associated to a
non-interacting system with Fermi velocity $v_F$ and is not differentiable, while
$R_{lm}(p)$, expressed by an all order expansion, is differentiable at $p=0$. 
By the lattice WI again we get relations between the bare
parameters
\be Z_0=Z\;,\qquad Z_1=Z_2=v_F Z\;.\label{thm2.1}\ee
so that
\be \hat \Pi_{lm}(p)= v_F^2
\media{\hat\jmath_{p,l};\hat\jmath_{-p,m}}_{0,v_F}+\hat
R_{lm}(p) \nn\ee
Note that $\hat \Pi_{lm}(p)$ is even and \bea && \s_{i}=-\frac{2}{3\sqrt3}\lim_{p_0\to
0^+}\frac1{p_0} \Big[ \big(\hat R_{ii}(p_0,\vec 0)- \hat
R_{ii}(0,\vec 0)\big)
\\
&&+\big(v_F^2 \media{\hat j_{(p_0,\vec 0),l};\hat j_{(-p_0,\vec
0),m}}_{0,v_F}- v_F^2 \media{\hat j_{\V0,l};\hat
j_{\V0,m}}_{0,v_F}\big)\Big]\;.\nn\eea
Note that $\hat R_{ii}$
 is an even function 
and has continuous derivatives; therefore the derivative vanishes at
$p=0$. In contrast the first term is identical to the non interacting
one but with a different Fermi velocity; but as
the free conductivity is independent from $v_F$ , the final result
is indeed universal and the following result is proved,
see \cite{57},\cite{58}
\vskip.3cm
\noindent
{\it 
For $|U|\le U_0$, if $U_0$ is a suitable constant 
\be \s_{lm}=
\frac{e^2}{h}\frac{\pi}{2} \d_{lm}\;. \nn\ee while the Fermi
velocity $v_F=3/2 t+b U+O(U^2)$.
}
\vskip.3cm
The above result is in agreement with the observed universality
of the optical conductivity in Graphene and with the increase of the velocity (for nearest neighbor interaction $b=0.3707...$.). It is essential to keep all the irrelevant terms due to the
lattice to preserve universality.

Another manifestation of universality in planar condensed matter systems is in the Hall effect,
appearing in systems with an emerging description in terms 
of massive Dirac fermions.
A combination of RG and Ward Identities analogue to the one described above leads 
to the proof that universality (quantization) persist even in presence of short-ranged many body interaction \cite{71},\cite{72}.

\subsection{Massless bosons and higher cut-offs} 

The above results have been obtained with
massive gauge fields or short range interactions and up
to cut-offs of the order of the inverse coupling. It is of
course important to extend to the massless
case and with larger cut-offs.

The model (6)  with $M=0$  
can be studied by a RG analysis decomposing both the
bosonic and fermionic fields. The independence from $\xi$
still holds from the validity of Ward Identities. In $d=3$
this analysis
was done in \cite{71} and it can be extended
to the model (6) with $M=0$. It leads to an expansion in terms
of a finite set of running coupling constants, whose coefficients
are finite and verify $n!$ bounds at order $n$. The consistency
of the method relies on the fact that the running
coupling constants are small for any $h$.
In $d=4$
in addition to $Z_h,Z^A_h$, $e_h$, $\n_h$, corresponding to the
fermionic and bosonic wave function renormalization,
the effective charge and the fermionic mass renormalization,
one obtains also non-gauge invariant running coupling
constants: a boson mass $2^{2h}  m_h$, $\k_h$ corresponding
to quartic boson terms and $R_h$ corresponding to
non transversal quadratic boson terms. The presence of
non gauge invariant couplings is related to the breaking
of gauge invariance in the intermediate RG steps due to the
momentum decomposition.

There is a basic difference in the analysis in the lattice
model (6) with $M= 0$ and in continuum QED model
with a momentum cut-off, as the one in 
\cite{23} - \cite{25} or \cite{27},\cite{28}. 
In the case
with momentum cut-off the WI are violated at finite cutoff,
and the flow of $m_h,\k_h,R_h$ is controlled introducing
counterterms in the bare action. This makes the momentum
regularization probably non suitable for a nonperturbative
approach as Ward Identities are recovered
only removing the ultraviolet cut-off. 
With a lattice cutoff,
instead, the Ward Identities are true with a finite lattice
step. Moreover, they can be used to control the flow
of the non gauge invariant running coupling.

The idea, see \S 5
and \cite{72} -\cite{74} 
is
to get information on the rcc 
introducing a reference model. In the case of (6) with $M=0$,
the reference model is (6) itself
but with boson propagator
\be
g^A_{\m,\n}(x,y)=
{1\over L^4}\sum_k \chi_h(k) { e^{i k (x-y)}\over |\s|^2+M^2}(\d_{\m,\n}-{\bar\s_\m \s_\n\over |\s|^2})\label{propo1}\ee
with $\chi_h(k)$ vanishing for $|k|\le 2^h$. The model (6)
and the reference model can be analyzed by a similar RG analysis
for scales $k\ge h$, and the rcc are the same. On the other hand in the
integration of the
 scales smaller than $h$ in the reference model one notes that
the boson fields disappear
and the running coupling constamts essentially
stop flowing. In addition, WI are true for the reference model, even
in presence of a momentum cut-off for the boson fields.
The correlations of the reference model are proportional to the rcc at scale $h$, hence the WI imply replations between the rcc at scale $h$.
One gets 
for the reference model
\be
\Pi_{\m\m}(p)=2^{2h }(m_h+R_\m)
\ee
with $R_\m=$
$O(\e_h^2)$ if $\e_h$
is the maximum of the rcc with scales greater than $h$. From the WI
$\sum_\m \s_\m\Pi_{\m\m}(p)=0$
one gets that $m_h=O(\e_h^2)$; 
that is the photon
mass $2^{2h }m_h$
stays bounded and vanishes as $h\to-\infty$
without the introduction of any counterterm. 
A similar argument can be repeated for 
the other non-gauge invariant couplings $\k_h, R_h$, while the flow of $\n_h$ is controlled by a suitable choice of the counterterm.
Finally
using the the vertex WI of the reference model we
get $\sqrt{Z^A_h}e_h=e (1+O(\e_h^2))$.
Therefore the WI have the effect that the flow of $e_h$ is
driven by 
$Z^A_h$ and by its flow equation
one obtains, $h=N,N-1,,$
\be
e^2_h={e^2\over 1-\log 2 {e^2\over 6\pi^2}(1+O(e^2))
h}
\ee
Therefore one expects that a control of the running coupling
constants if $Ne^2$ is smaller than some constant is obtained, that
is up to exponentially high cut-off. This analysis
is perturbative in the renormalized expansion; a non perturbative
result requires a further decomposition of the boson fieds $A_\m$ in
small and large regions, see e.g. \cite{75}, but in any case the
control of the flow of the rcc is an essential prerequisite.

When the boson is massive $M\not=0$, the second term of
the boson propagator is not decaying. However using the
$\x$-independence one can choose
$\x=0$ so that the boson propagator is ${\d_{\m\n}\over k^2+M^2}$. One
can distinguish two regions in the multiscale integration,
separated by a scale $2^{\bar h}=M$: for scales greater than $\bar h$ 
the theory has a renormalizable behavior and one
can repeat the same analysis sketched above for the $M=0$
case, while in the second region the boson can be integrated
out and the theory reduces to a fermionic theory
quartic in the fermions described in \S 2, that is has a non-renormalizable behavior. One therefore expects
that an exponentially high cut-off can be reached
in the massive case $M\not=0$. 
The argument
at the basis of the anomaly non-renormalization seen
above in \S 2.6 cannot however be applied with such exponentially high cut-off; the decomposition (61)
of the current correlation in the non interacting part
(with renormalized parameters) and a differentiable contribution
is not anymore true, and similarly the decomposition
(56) of the 2-point function in the free part (with
renormalized parameters) and a subleading correction.
The above decomposition are indeed peculiar to a nonrenormalizable
theory. There are however examples in which the anomaly non-renormalization holds also in renormalizable
cases, as in $d=2$ model, as discussed below.

There are similarly open problems on the role of
long range interactions on universality properties in condensed
matter. For instance, in the case of Graphene the
results in \S 3.3 ensures the universality in the optical conductivity
in the case of short range interactions, but in several
situations Graphene can be described with long range
forces. In this case the analysis in \cite{71} 
, extending a previous
study in the continuum in \cite{76}, shows that the rcc corresponding
to the fermion-boson interaction is essentially not flowing $e_h\sim e$ in contrast with (84).
A lowest order computation
of the conductivity is in \cite{77}, where universality
is still recovered, while in \cite{78} a $1/N$ expansion is performed finding corrections which are nevertheless still universal; a 
non-perturbative conclusions has not been still reached.
Similarly it would
be interesting to consider long range interactions in Weyl
semimetals or Hall insulators.

\section{Anomaly cancellation in a chiral theory}

\subsection{A chiral lattice $U(1)$ Gauge theory}

One of the main application
of the non-renormalization of the anomalies is
in the anomaly cancellation in a chiral gauge theory, like
the Standard Model.  
We consider a lattice chiral gauge theory, 
given by $2N$ massless fermions in four dimensions, labeled by an index $i=1,...,2N$;
we also define the indices $i_1=1,...,N$ and $i_2=N+1,...,2N$.
As before the correlations are obtained as derivatives of the generating function 
\be
e^{W(J,J^5,\phi)}= \int P(d A) \int P(d\psi) e^{V(\psi, A, J)+V_c(\psi)+
B(J^5, \psi)+(\psi,\phi)}
\label{llll}
\ee
where the bosonic integration has propagator (12) and the fermionic 
propagator has propagator  (10). If $V=V_1+V_2$ we call $O^+_{\m,i,s,x}={1\over 2 }\psi^+_{i,s, x}\s_\m^s   \psi^-_{i,s, x+e_\m a}$ and
$O^-_{\m,i,s,x}
=$
$
-{1\over 2 }\psi^+_{i,s, x+e_\m a}\s^s_\m \psi^-_{i,s, x}$, $s=L,R$, and  we 
define
\be
V_1(A,\psi,J)=
a^4\sum_{i,s,x}  [O^+_{\m,i,s,x} G_{\m,i,s,x}^+ +
O^-_{\m,i,s,x} G_{\m,i,s}^- ]
\ee
with $G^{\pm}_{\m,i,s}(x)=a^{-1}  (e^{\mp i a  Y_i(\l  b_{i,s} A_{\m,x}+ J_{\m,x})}-1)$ and $b_{i_1,L}=b_{i_2,R}=1$, $b_{i_1,R}= b_{i_2,L}=0$. If $J_\m=0$ $V_1$ represents the interaction; note that only the $L$
chirality of the $i_1$ fermions and the $R$ chirality of the $i_2$ fields
 interact with the boson field $A_\m$. Moreover
\bea
&&V_2(A,\psi,J)={r\over 2} a^4\sum_{i,x} [ \psi^+_{i,L, x}H_{\m,i,x}^+
\psi^-_{i,R, x+e_\m a}
+\nn\\
&&\psi^+_{i,L, x+e_\m a} H_{\m,i,x}^-\psi^-_{i,R, x}
 +\nn\\
&&\psi^+_{i,R, x}H_{\m,i,x}^+\psi^-_{i,L, x+e_\m a}+\psi^+_{i,R, x+e_\m a} 
H_{\m,i,x}^-\psi^-_{i,L, x}]
\eea
with $H_{\m,i,x}^\pm=
a^{-1}  (e^{\mp i a Y_i J_{\m,x}}-1)$.
The mass counterterm is
$
V_c=  \sum_{i} a^{-1}  \n_i  a^4\sum_x (\psi^+_{i,L,x}\psi^-_{i,R,x}+\psi^+_{i,R,x}\psi^-_{i,L,x})
$
and \be
B=a^4 \sum_{\m,x} J^5_{\m,x} j^5_{\m,x} \quad j^5_{\m,x}=\sum_{i,s} \tilde \e_i \e_s Y_j Z_{i,s}^5 \psi^+_{x,i,s}\s_\m^s \psi^+_{x,i,s}
\ee
with $\tilde \e_{i_1}=-\tilde \e_{i_2}=1$ and $\e_{L}=-\e_{R}=1$. 

The fermionic 2-point function is
$S_{i,s,s'}(x,y)$ is the derivative with respect to 
$\phi^+_{i,s,x}$, $\phi^-_{i,s',y}$ of $W$,  the vertex function
$\G_{\m, i, s}(z,x,y)$ 
is the derivative respect to
$J_{\m, z},\phi^+_{i,s,y}
\phi^-_{i,s,y}$ and the chiral vertex  
$\G^5_{\m, i, s}(z,x,y)$ 
is the derivative respect to
$J^5_{\m, z},\phi^+_{i,s,y}
\phi^-_{i,s,y}$.
The three current 
vector $VVV$ and axial $AVV$ correlations are
\be \Pi^{VVV}_{\m,\n, \r}(z,y,x)=
{\partial^3 \WW_\L\over \partial J_{\m,z}
\partial J_{\n,y}\partial J_{\r,x}}|_0 \ee
and $
\Pi^{AVV}_{\m,\n, \r}(z,y,x)={\partial^3 \WW\over \partial J^5_{\m,z}
\partial J_{\n,y}\partial J_{\r,x}}|_0$. Again by performing the change of variables
$
 \label{wis}\psi^\pm_{i,s,x}\to
\psi^\pm_{i,s,x} e^{\pm i Y_i \a_{x} }$
we get
\be
W(J,J^5,\phi)=
W(J+d_\m \a ,J^5, e^{i Y \a}
 \phi)\ee
where $J+d_\m \a$ is a shorthand for $J_{\m,x}+d_\m \a_{x}$
and $e^{i Y\a}\phi$ is a shorthand for $e^{\pm i Y_i\a_{x}} \phi_{i,s, x}^{\pm}$; by differentiating we get the WI
\bea
&&\sum_{\m_1} \s_{\m_1}(p_1+..p_{n-1})
\hat \Pi^{VVV}_{\m_1,..,\m_n}(p_1,.,p_{n-1})=0\nn\\
&&
\sum_\m \s_\m(p) \hat\G^\L_{\m,i,s}(k,p) 
= Y_i
(\hat S_{i,s,s}(k)-\hat S_{i,s,s}(k+p))\label{wia}\\
&&
\sum_\n \s_\n(p_1)
\hat \Pi^{AVV}_{\m,\n,\r}(p_1,p_2)=\sum_\r \s_\r(p_2)
\hat \Pi^{AVV}_{\m,\n, \r}(p_1,p_2)=0\nn
\eea
The mass counterterms $\n_i$ has to be chosen so that the
2-point correlations are singular at $k=0$, that is the
fermionic dressed mass is vanishing and 
the renormalizations
$\ZZ^5_{i,s}$
is chosen to ensure that
\be
\lim_{k,p\to 0}{\hat\G^5_{\m, i,s}(p,k)\over\hat\G^5_{\m, i,s}(p,k)}=\e_s
\ee
ensuring that the vector and axial part of the current of
each particle carry the same charge. The total current coupled to $A_\m$ is 
\be
j^T_\m=\sum_{i_1} Y_{i_1}
\psi^+_{i_1, L,x} \s_\m^L
\psi^-_{i_1, L,x}+\sum_{i_2} Y_{i_2}
\psi^+_{i_2, R,x} \s_\m^R 
\psi^-_{i_2, R,x}
\ee
and the axial and vector part of the current is
\be
j^{T,V}_\m={1\over 2}\sum_{i} Y_i j_{\m,i,x}\quad\quad j^{T,A}_\m={1\over 2}\sum_{i} Y_i \tilde\e_i j^5_{\m,i,x}\label{prtr}
\ee
with $\tilde\e_{i_1}=-\tilde\e_{i_2}=1$,
$j_{\m,i,x}=\bar\psi_{i,x}\g_\m\psi_{i,x}$, $ j^5_{\m,i,x}=\bar\psi_{i,x}\g_5\g_\m\psi_{i,x}$
and $\psi_{i,x}=(\psi^-_{i,L,x},\psi^-_{i,R,x})$, $\bar\psi_{i,x}=(\psi^+_{i,R,x},\psi^+_{i,L,x})$.
Note the chiral nature of the theory, as in the current the fermion with different chirality have different charges.

In the formal continuum limit the action reduces to 
$\int dx \{
F_{\m,\n}F_{\m,\n} +$
\bea
&&\sum_{i_1} [\psi^+_{i_1, L,x} \s_\m^L (\partial_\m+\l  Y_{i_1} 
 A_\m)
\psi^-_{i_1, L,x}+\psi^+_{i_1, R,x} \s_\m^R \partial_\m
\psi^-_{i_1, R,x}]
\\
&&\sum_{i_2} [
\psi^+_{i_2, R,x} \s_\m^R (\partial_\m+\l  Y_{i_2} 
 A_\m)
\psi^-_{i_2, R,x}+\psi^+_{i_2, L,x} \s_\m^L \partial_\m
\psi^-_{i_2, L,x}]\}
\label{h111}\nn\eea
Note that the $R$ fermions of kind $i_1$ and 
the $L$ fermions of kind $i_2$ decouple and
are
fictitious, non interacting degrees of freedom, which are convenient to introduce in view of the lattice 
regularization \cite{B3c}.
If $N_1=N_2=4$ (85) is a lattice regularization of the U(1)
sector of the Standard Model with no Higgs and massless
fermions; in this case
$i_1=(\n_1,e_1,u_1,d_1)$ are the left handed components and $i_2=(\n_2,e_2,u_2,d_2)$ the right handed of the leptons and quarks.

\subsection{Anomaly cancellation}
In the non-interacting case the anomaly is the sum of the
anomalies of the particles weighted by $\tilde\e_i Y_{i,s}$, that is
up to $O(a^\th |p|^{2+\th})$ terms
\be
-i p_\m \Pi^{AVV,0}_{\m,\r,\s}={1\over 2\pi^2}
\e_{\a, \b,\r,\s} p^1_\a  p^2_\b[\sum_{i_1} Y^3_{i_1}-\sum_{i_2} Y^3_{i_2}]\ee
which, under the anomaly cancellation condition 
$[\sum_{i_1} Y^3_{i_1}-\sum_{i_2} Y^3_{i_2}]=0$
is vanishing (p to subleading terms. In the
case of the U(1) sector of the StandardModel, this condition
is verified. Indeed the physical values 
$Y_{\n_1}=Y_{e_1}=-1$, $Y_{u_1}=Y_{d_1}=1/3$,
$Y_{\n_2}=0$, $Y_{e_2}=-2$, $Y_{u_2}=4/3$, $Y_{d_2}=-2/3$
corresponding to the electric charges $e(0,-1,2/3,-1/3))$
verify the condition.

The issue is if the anomaly cancels under the same
condition in presence of interaction. Higher order corrections
could require other conditions to vanish, impossible
to verify. The interacting theory can be analyzed
via a multiscale analysis similar to the one for the vector
model in \S 2. One first integrate the $A_\m$ field and then perform
a multiscale analysis for the Grassmann variables:
after integrating $\psi^N,...,\psi^{h+1}$
one gets
\be
e^{W(J,J^5)}=\int P_{Z_h}(d\psi^{(\le h)})
e^{
V^{(h)} ( \sqrt{Z_h} \psi^{(\le h)},J,J^5)}\label{dds1}
\ee
with $P_{Z_h}(d\psi^{(\le h)})$ with propagator
$
\hat g_{i}^{(\le h)}(k)=$
\be
\chi_{h}(k)
(\sum_{\m} \g_0\tilde\g^{h}_\m a^{-1}  i \sin (k_\m a)+  
a^{-1}\hat\g^{h}_0 \sum_\m (1-\cos k_\m a))^{-1}\label{sapa1}
\ee
with
\be
\tilde \g^{h}_0= \begin{pmatrix} 0& Z_{h,L,i} I  \\  Z_{h,R,i}I & 0  \end{pmatrix}\quad \tilde \g^h_j=\begin{pmatrix} 0& i Z_{h,L,i} \s_j \\-i Z_{h,R,i}\s_j& 0 \end{pmatrix}
\ee
and $V^{(h)}(\sqrt{Z_h}\psi,J,J^5)=R V^h+$
\bea
&&a^4 \sum_x \sum_{i,s}  [\n_{h,s} 2^{h}
\sqrt {Z_{h,L,i}Z_{h,R,i}}  
(\psi^+_{i,L,x}\psi^-_{i,R,x}+\psi^+_{i,R,x}\psi^-_{i,L,x})+\nn\\
&&+
Z_{i,s,h}^{J}
J_{\m,x} \psi^+_{i,s,x}\s_\m^s \psi^-_{i',s,x}+\e_s \tilde\e_i 
Z_{i,s,h}^{5} J^5_{\m,x}
\psi^+_{i,s,x}\s_\m^s \psi^-_{i,s,x}]\label{ess11a}
\eea
and $R V^h$ is sum of monomials of $l$ fields with  $4-3/2l-q -m<0$. The
interaction produces a different wave function renormalizations
depending on the type of particles and on the chirality;
again in the limit $h\to\io$ one gets for a proper $\n_i$ that $\n_{i,h}\to 0$,
$Z_{i,s,h}\to Z_{i,s}$, $Z^J_{i,s,h}\to Z^J_{i,s}$,  $Z^5_{i,s,h}\to Z^5_{i,s}$
with $Z_{i,s}, Z^J_{i,s}, Z^5_{i,s}$
depending on $i , s$ and all the
lattice details. In particular the wave function and the vertex renormalizations depend on the type of the particle and from its chiraity.

The current correlations can be written as
\be
\hat \Pi^{AVV}_{\m,\r,\s}(p_1,p_2)=\hat \Pi^{AVV,a}_{\m,\r,\s}(p_1,p_2)+\hat\Pi^{AVV,b}_{\m,\r,\s}(p_1,p_2)
\ee
where $\hat\Pi^{AVV,b}_{\m,\r,\s}(p_1,p_2)$ with continuous derivatives and
$\hat \Pi^{AVV,a}_{\m,\r,\s}$ containing only the dominant part of the propagator and marginal terms
\bea
&&\sum_{ji,s}\sum_{h_1\atop h_2,h_3}\tilde\e_{j}\e_s Y^3_{j} { Z^{5}_{h_1,j,s}\over Z_{H_1,j,s}} {Z^{J}_{h_2, j,s}\over Z_{h_2,j,s}}
{Z^{J}_{h_3, j, s}\over Z_{h_3,j,s}}\label{fon11} \\
&&\int {dk \over (2\pi)^4}{\rm Tr}{f_ {h_1}(k)\over i  \s^{s}_\m k_\m}
i\s^{s}_\m{ f_{h_2}\over i \s^{s}_\m (k_\m+p_\m)} i \s^{s}_\n {f_ {h_3}\over i   \s^{s}_\m (k_\m+p^2_\m)}(i \s^{s}_\r)\nn
\eea
Note the presence of the factors 
${ Z^{5}_{h_1,j,s}\over Z_{h_1,j,s}}$ and  ${Z^{J}_{h, j,s}\over Z_{h,j,s}}$ depending on the particle amd chiral index.
The two-point and vertex
correlations are given by
$\hat S_{i,s,s}(k)={1\over Z_{i,s} -i \s_\m^s k_\m}(1+r_1(k))$, 
$\G_{\m,i,s}
(k,p)=$
\be
{Z Z^J_{i,s}\over Z_{i,s}^2} {1\over -i \s_\m^s k_\m}i \s_\m^s {1\over -i \s_\m^s (k_\m+p_\m)}[1+r_2(p,k)]
\ee
with $r_1=O((a |k|)^\th)$ and $r_2=O(a^\th(|k|^\th+|k+p||^\th)$
we get \be
{Z^{J}_{i,s}\over Z_{i,s}}=1\ee
By choosing $\ZZ^5_{i,s}$ imposing
$Z^J_{i,s}/Z^5_{i,s}=1$
 and using that the limit is reached exponentially fast we get
\be
\hat \Pi^{AVV}_{\m,\r,\s}(p_1,p_2)=(\sum_i \tilde\e_i Y^3_i)\hat \Pi^{5,0,a}_{\m,\r,\s}(p_1,p_2)
+\hat R^{AVV,b}_{\m,\r,\s}(p_1,p_2)
\ee
with $\hat R^{AVV,b}_{\m,\r,\s}(p_1,p_2)$ with continuous derivatives. Proceeding as in \S 2.6 using the WI (91) one proves therefore the following, see \cite{55},\cite{79},
\cite{80}.
\vskip.3cm
\noindent
{\it For $|\l|\le \l_0 (Ma)$, it is possible to find $\n_i$,  $\ZZ_{i,s}^5$ continuous functions in $\l$ such that
the AVV correlation verifies
\bea
&&\sum_\m \s_\m(p_1+p_2) \hat\Pi^{5}_{\m,\r,\s}(p_1,p_2)=
\sum_{\m,\n} \e_{\a,\b,\r,\s}\nn\\
&&
{1\over 2 \pi^2}p^1_\a p^2_\b [\sum_{i_1} Y_{i_1}^3-\sum_{i_2} Y_{i_2}^3] +r_{\r,\s}(p_1,p_2)\label{1} 
\eea
with $|r(p_1,p_2)|\le C a^\th \bar p^{2+\th}$, 
$\bar p=\max (|p_1|, |p_2|)$.}
\vskip.3cm
The vanishing of the anomaly holds therefore up
to cut-off of the order of the inverse coupling under
the same condition as in the continuum, up to subleading corrections.

In order to increase the size of the cut-off one would
like to proceed as in \S 3.3, and for massive bosons one
needs the $\x$ invariance to ensure that the non decaying part of the gauge propagator is not contributing, so that there is a reduction of the degree of divergence. However such invariance is based 
on the WI for the current 
associated to the gauge field. This requires two sets of WI for the total and chiral current, but the second are violated by anomalies for generic values of the hypecharges.
The above result shows that, at least for cut-off of the order of the inverse coupling the WI
for the chiral current is preserved, at least for the $AVV$ correlation
which is
the dominant one in the continuum limit. 
This is a prerequisite
condition for the construction of the $U(1)$ sector of the Standard Models up to exponentially
high cut-off, as an extension of this result should possibly ensure
that the contribution of the non decaying
part of the boson is vanishing or at least small at higher energy scales.

\section{Massive $U(1)$ gauge theory in $d=2$} 

\subsection{The lattice Sommerfield model} 

Let us consider now what happens in the $d=2$ vector
model (6), see \cite{81}. In this case the dimension
is $D=2-l/2-m$ corresponding to a renormalizable degree of divergence; however, if the contribution of the non decaying part of the boson propagator vanishes  one
passes from a  renormalizable to a superrenormalizable degree of divergence (as in $d=4$ there is a reduction from a non-renormalizable to
a renormalizable behavior). In $d=2$, however, the theory can be constructed for values of the cut-off arbitrarily large. The reduction
of the degree of divergence appears in the fact that the bare parameters can be chosen independent on the cut-off; in absence of such reduction
the theory is essentially equivalent to the Thirring model  
and
the bare wave function renormalization would vanish with the cut-off.
Note that such a reduction does
not appear in the exact solution of the continuum version
of this model \cite{82} as a momentum regularization is
used violating WIs.

Using a lattice cut-off as in (6) the $\x$ independence allows to choose $\x=0$. 
After the integration of the $A_\m$ field, one obtain a
purely fermionic theory with a short range interaction
with range $O(1/M)$. One has to distinguish two regimes
distinguished by a scale $2^{\bar h}=M$ . In the integration of the
scales higher than $\bar h$ there is an improvement in the bounds
due to the non locality of the interaction, similar to
the one happening in the non-local Thirring model 
\cite{72}-\cite{74}, \cite{83}.

\insertplot{220}{110}
{\ins{10pt}{70pt}{$H^h_{2,0}$}
\ins{50pt}{65pt}{$=$}
\ins{75pt}{90pt}{$H^h_{0,1}$}
\ins{100pt}{45pt}{$H^h_{2,0}$}
\ins{130pt}{65pt}{$H^h_{2,0}$}
\ins{180pt}{80pt}{$H^h_{2,1}$}
\ins{110pt}{65pt}{$+$}
\ins{160pt}{65pt}{$+$}
}
{figjsp467aa11}
{\label{h2} Decomposition of $H^h_{2,0}$
} {0}

The kernels with 2 or 4
fermionic fields are dimensionally marginal or relevant; however
a suitable decomposition of the kernels $H^h_{l,m}$, see Fig.5,
allows
to improve their scaling dimension using the non
locality of the interaction, see \cite{81}. This implies the irrelevance
of all effective interactions
in the RG sense and establishes
the reduction of the degree of freedom; the ultraviolet
cut-off can be removed with finite bare couplings.

In the integration of the scales smaller than $\bar h$ the non irrelevant
part of the potential has the form $\l \int dx dy v(x-y) j_\m j_\m$; in contrast with the $d=4$ case the
quartic coupling constant is marginal. However the theory
can be still controlled thanks to the vanishing of the
beta function, proved in \cite{72}.

The fact that in the second regime there are quartiic
marginal interactions causes a striking difference in the
chiral correlations with respect to the $d=4$ case. The
chiral current correlation can be written as
\be
\hat\Pi^5_{\m\n}=\hat \Pi^{5,a}_{\m\n}+\hat R_{\m\n}\label{lal}
\ee
where $\hat \Pi^{5,a}_{\m\n}$ contains no irrelevant terms
and contains only the  dominant part of the propagator (the first term in (27)) and  $\hat R_{\m\n}$ is the rest. As in the non-interacting
case $\hat \Pi^{5,a}_{\m\n}$
is non continuous in p while $\hat R_{\m\n}$ is continous.
There is however a crucial difference with respect to the $d=4$ case;
in that case the only marginal terms were bilinear in the $\psi$
and $\hat \Pi^{5,a}_{\m\n}$ was only given by sum of triangle graphs.
In $d=2$., in contrast, the marginal terms are quartic interactions and
$\hat \Pi^{5,a}_{\m\n}$ is expressed by an infinite series of terms.

One cannot therefore compute explicitly $\hat \Pi^{5,a}_{\m\n}$
but has to follow a different strategy. One
introduces a reference model
\bea
&&e^{W_T(J^+,J^-,\phi)}=\int  P(d\psi^{(\le N)})\\
&&e^{
\tilde\l  \tilde Z^2 \tilde\l \int dx dy v(x-y) \r_{+,x}\r_{-,y}
+\sum_\o[
\tilde Z^+ \int dx J^+_{\o,x} \r_{\o,x}+\tilde Z^- \int dx J^-_{x} \o \r_{\o,x}]
}\label{ddd1}\nn
\eea
with $\o=\pm$, $x\in \RR^2$ is a continuum variable,
$P(d\psi)$ is the Grassmann integration with propagator 
${1\over \tilde Z}{\chi_N\over -i \not k}$, $\chi_N(k)$ smooth cut-off function non vanishing in  $|k|\le 2^N$,
 $v(x - y)$ 
a short range symmetric potential, $\hat v(0)=1$, $\r_\o=\psi^+_\o\psi^-_\o$. The model can be considered as the scaling limit of the model (6) in $d=2$ with a momentum cut-off. A similar RG analysis can be done also for this model \cite{72}-\cite{74}, showing
that there exists a suitable choice of its parameters, actually $\tilde Z^\pm, \tilde Z, \tilde \l$, such that the corresponding running coupling constants tend as $h\to-\io$ 
to the same limiting value than in the lattice model (6).
This implies that the 2-point and the vertex functions of the two models coincide up to subdominant terms in the momenta, and the current correlations coincide up to terms continuos in $p$.
The reference model is defined in the continuum with a momentum cut-off;
in contrast with previous lattice models, it verifies
the local chiral symmetry $\psi^\pm_{\o,x}\to e^{\pm i \a_\o} \psi^\pm_{\o,x}$.
\subsection{Anomaly non-renormalzation}

We can indeed derive Ward Identies also for the model (108) by the transformation
$\psi^\pm_\o\to e^{\pm i \a_{\o,x} }\psi^\pm_{\o,x}$. 
One obtains
\bea
&&D_\o <\hat\r_{p,\o}\hat\psi^+_{k,\o'}\hat\psi^-_{k+p,\o'}>
+
\D_N(k,p)=\nn\\
&&
\d_{\o,\o'}  {1\over \tilde Z} [<\hat \psi^+_{k,\o'} \hat \psi^-_{k,\o'}>  - <\hat\psi^+_{k+p,\o'} 
\hat\psi^-_{k+p,\o'}>\label{lll}
\eea
where $D_\o=-ip_o+\o p$, $\o=\pm$, 
\be
\D_N=<\d\hat\r_{p,\o}\hat\psi^+_{k,\o'}\hat\psi^-_{k+p,\o'}
>\ee
and
$
\d\hat\r_{p,\o}=
\int d k$
\be
[(\chi^{-1}_N(k+p)-1)D_\o(k+p)
-(\chi^{-1}_N(k)-1)D_\o(k)]\hat\psi^+_{k,\o}\hat\psi^-_{k+p,\o}\nn\ee

The momentum cut-off produces the extra
term $\D_N$ in the WI for the current and the current
, which are both violated. The above identity can be also derived in perturbation theory by (31).

The
correction term $\D_N$, which would be not present neglecting
the cut-off, is expressed by a complicate perturbative expansion;
however using a detailed decomposition of the
correction term, 
\cite{72}-\cite{74}, \cite{83}, in the limit $N\to\io$
one gets
\be
\lim_{N\to\io} \D_N(k,p)=\t\hat v(p)  D_{-\o}(p)
<\hat\r_{p,-\o}\hat\psi^+_{k,\o'}\hat\psi^-_{k+p,\o'}>\label{sapp}\nn
\ee 
with $\t={\tilde\l \over 4\pi }$. Note that $\t$ is linear in $\tilde\l$; all higher order corrections cancel. In the limit $N\to\io$ the following WI holds for the model \pref{ddd1}
\be
-i (1- \t \hat v(p)) p_\m \tilde \G_{\m}(k,p)={\tilde Z^+\over \tilde Z}
(\tilde S(k)-\tilde S(k+p))\label{spp1}\ee
and
$i p_\m \tilde \Pi^5_{\m,\n}={\tilde Z^+ \tilde Z^-\over 2\pi \tilde Z^2}
{\e_{\m\n}p_\m\over (1+\t \hat v(p))  }$
\be i p_\n \tilde \Pi^5_{\m,\n}={\tilde Z^+ \tilde Z^-\over 2 \pi \tilde Z^2}
{\e_{\n\m}p_\n\over  (1-\t \hat v(p))  }\label{ss1}
\ee
where  $\ZZ^5\tilde\Pi^5_{\m,\n}$ is equal to $\hat \Pi^{5,a}_{\m,\n}$ in (107).

The parameters $\tilde Z^\pm,\tilde \l,\tilde Z$ are non trivial unknown functions,
depending on all the details of the regularizations. However
the $\tilde Z, \tilde\l, \tilde Z^\pm$ are chosen so that the vertex and 2-point functions of the lattice and continuum model are the same; therefore
the lattice WI (19) and the WI obtained in the reference model
\pref{ddd1}  must be the same and this implies a relation between such parameters
\be{ \tilde Z^+ \over \tilde Z(1-\t)}=1\ee 
In addition from the definition of $\ZZ^5$
\be
{ \tilde Z^+ \over \tilde Z(1-\t)}=\ZZ_5 { \tilde Z^- \over \tilde Z(1+\t)}=1\ee
from which $\ZZ_5=(1+\t){ \tilde Z\over \tilde Z^-}$ and, up to $O(p^2)$ terms
\be
i p_\m \tilde \Pi^5_{\m,\n}={(1-\t)\over 2\pi \ZZ^5}
\e_{\m\n}p_\m  \quad\quad  
i p_\n \tilde \Pi^5_{\m,\n}={(1+\t)\over 2 \pi \ZZ^5}
\e_{\n\m}p_\n\label{ssa1}\ee
We use now the above expressions in \pref{lal}.
The WI imply
\be
i p_\n \hat\Pi^5_{\m,\n}(p)={(1+\t)\over 2 \pi }
\e_{\n\m}p_\n
+p_\n R_{\m,\n}(p)=0\nn
\ee
so that $
\hat R_{\m,\n}(0)=-(1+\t) \e_{\n\m}/2\pi $
where we have used that
$\hat R_{\m,\n}(p)$
is continuous. In conclusion
\bea
&&i p_\m\hat\Pi^5_{\m,\n}(p)= {1\over 2\pi}p_\m
 [\tilde\Pi^{5,a}_{\m,\n}(p)+\hat R_{\m,\n}(p)]=\nn\\
&&[(1-\t)\e_{\m,\n}-(1+\t)\e_{\n,\m}] p_\m/2\pi=1/\pi\e_{\m,\n}p_\n\nn
\eea
that is all the dependence of the coupling disappears in the anomaly, see \cite{82}
\vskip.2cm
\noindent
{\it If $e^2\le  
e_0$ with $e_0$ independent on $L, a$, for a suitable $\ZZ^5$,
 the correlations are given by convergent expansions and, in the $a\to 0$ limit
\be i p_\m \hat\Pi^5_{\m,\n}={1\over \pi}
\e_{\n,\m} p_\m  +O(|p|^{1+\th})\ee}

The uv cut-off can be removed in $d=2$ $U(1)$ model with finite bare parameters in the continuum $a\to 0$ limit.
The fact that the bare parameters are finite follows from the $\x$-independence, and it is a consequence of the reduction
of divergence degree from a renormalizable to a superrenormalizable one.
The arguments at the basis of the anomaly non renormalization proof in $d=4$
seen in \S 2 does not hold: the 2-point and vertex functions
in presence of interaction are not equal to the free ones with renormalized parameters. Despite this fact
the anomaly renormalization holds.

The above result is based on the validity of the Ward Identities in the lattice model. One could consider also the chiral model (85)
in $d=2$. In that case the anomaly in the non-interacting case
cancels under the condition $\sum_{i_1} Y_{i_1}^2-\sum_{i_2} Y_{i_2}^2$. It is therefore a natural question if the anomaly cancels also with such a condition in presence of interaction, and if the cancellation in the chiral WI would ensure the reduction of the degree of divergence
and the possibility of taking the continuum limit $a\to 0$ with finite bare parameters,
as in the Sommerfield model. This would provide the analogue 
of the construction of the $U(1)$ sector of the Standard Model with large cut-off at a non perturbative level.

\subsection{Luttinger liquids}

One dimensional metals have typically an emerging description
in terms of massless Dirac fermions with a quartic ferminic interaction. 
In this case there is a different form of universality; transport coefficients 
are function of the microscopic parameter but they verify universal Luttinger liquid relations. While such relations can be checked in special solvable models, they indeed are valid in a wide class of non solvable models;
in particular the following relation 
between Drude weight $D$, susceptibility $\k$ and Fermi velocity $v_F$can be proved:
\begin{equation}\label{eq:Hrel}
\frac{D}{\kappa} = v^{2}\;.
\end{equation}
The validity of this identity
was proved in
\cite{83}-\cite{87} using a similar strategy as in \S 5.1, 5.2;
one performs an RG analysis of the non relativistic
 model and use the properties of the reference model (107).
The validity of such relations is strictly connected to the non-renormalization of $\t$ present in the WI of the reference model.

Similar ideas have been used to establish the bulk-edge correspondence in Hall insulators  in presence of interaction \cite{88}. Even when 
at the edge of Hall insulators there are fermions with both chiralities
which interact with short range potential, the conductivity is non-renormalized.

\section{Outlook}

We have reviewed a new approach allowing to establish
non-renormalization properties of anomalies.
In contrast with previous studies, the lattice terms are fully taken into account and the results are non-perturbative, being based 
on convergent expansions
By such methods it was possible to establish the anomaly non-renormalization in vector U(1) models and the anomaly cancellation
in chiral U(1) models with lattices of the order of the inverse coupling
in $d=3+1$, and the non-renormalization for any lattice in $d=1+1$.
They also allowed to prove universality in transport coefficients in several 
materials with short range interactions, including Weyl semimetals, graphene and Luttinger liquids. 

The cancellation of the anomalies with a finite lattice is
a natural starting point for 
the non-perturbative construction of chiral gauge models 
like the electroweak theory with high cut-off. Similarly the short range interaction is 
a starting point for understanding the universality in transport with long range Coulomb interaction in graphene or Hall insulators. 

\vskip.2cm
{\bf Aknowledgments} I am indebted with A. Giuliani and M.Porta, 
my collaborators in a large part of the research reported here.
The work supported 
by MUR (Italian Ministery of research), Grant
No. PRIN201719VMAST01 and the GNFM.

\end{document}